\documentclass[a4paper,fleqn,usenatbib]{mnras}

\usepackage[T1]{fontenc}
\usepackage{ae,aecompl}

\usepackage{graphicx}	
\usepackage{amsmath}	
\usepackage{amssymb}	
\usepackage{rotating}
\usepackage{listings}
\usepackage{threeparttable}
\usepackage{hyperref}
\usepackage{newtxtext,newtxmath}

\title[IGR\,J18249--3243: a new GeV-emitting FR\,II] {IGR\,J18249--3243: a new GeV-emitting FR\,II and the emerging population of high energy radio galaxies}

\author[G. Bruni et al.]
{G. Bruni$^{1}$\thanks{Contact e-mail:\href{mailto:gabriele.bruni@inaf.it}
{gabriele.bruni@inaf.it}},
 L. Bassani$^{2}$,
 M. Persic$^{3,4}$, 
 Y. Rephaeli$^{5,6}$, 
 A. Malizia$^{2}$, 
 M. Molina$^{7}$, 
 M. Fiocchi$^{1}$, 
 R. Ricci$^{8}$,
 \newauthor 
 M. H. Wieringa$^{9}$, 
 M. Giroletti$^{8}$, 
 F. Panessa$^{1}$, 
 A. Bazzano$^{1}$, 
 P. Ubertini$^{1}$\\
$^{1}$ INAF -- Istituto di Astrofisica e Planetologia Spaziali, via Fosso del Cavaliere 100, I-00133 Roma, Italy\\
$^{2}$ INAF -- Osservatorio di Astrofisica e Scienza dello Spazio, via P. Gobetti 101, I-40129 Bologna, Italy \\
$^{3}$ INAF -- Osservatorio Astronomico di Trieste, via G.B. Tiepolo 11, I-34100 Trieste, Italy \\
$^{4}$ INFN -- Trieste, via A. Valerio 2, I-34127 Trieste, Italy \\
$^{5}$ School of Physics and Astronomy, Tel Aviv University, Tel Aviv 69978, Israel \\
$^{6}$ Center for Astrophysics and Space Sciences, University of California at San Diego, La Jolla, CA 92093, USA \\
$^{7}$ INAF -- Istituto di Astrofisica Spaziale e Fisica Cosmica, Via A. Corti 12, I-20133 Milano, Italy \\
$^{8}$ INAF -- Istituto di Radioastronomia, via P. Gobetti 101, I-40129 Bologna, Italy \\
$^{9}$ CSIRO -- Astronomy and Space Science, PO Box 76, Epping, New South Wales 1710, Australia 
}

\date{Accepted XXX. Received YYY; in original form ZZZ}

\pubyear{}

\begin{document}
\label{firstpage}
\pagerange{\pageref{firstpage}--\pageref{lastpage}}
\maketitle

\begin{abstract}
The advent of new all-sky radio surveys such as the VLA Sky Survey (VLASS) and the Rapid ASKAP Continuum Survey (RACS), performed with the latest generation radio telescopes, is opening new possibilities on the classification and study of extragalactic $\gamma$-ray sources, specially the underrepresented ones like radio galaxies. In particular, the enhanced sensitivity (sub-mJy level) and resolution (a few arcsec) provides a better morphological and spectral classification. In this work, we present the reclassification of a \emph{Fermi}-LAT source as a new FRII radio galaxy from the \emph{INTEGRAL} sample found to emit at GeV energies. Through a broad-band spectral fitting from radio to $\gamma$-ray, we find that the commonly invoked jet contribution is not sufficient to account for the observed $\gamma$-ray flux. Our modeling suggests that the observed emission could mainly originate in the lobes (rather than in the radio core) by inverse Compton scattering of radio-emitting electrons off the ambient photon fields. In addition, we cross-correlated the latest generation radio surveys with a list of \emph{Fermi}-LAT objects from the literature considered to be candidate misaligned AGN, finding four new radio galaxies with a double-lobed morphology. Additional four objects could be classified as such thanks to previous studies in the literature, for a total of nine new radio galaxies with GeV emission presented in this work. We foresee that further objects of this class might be found in the near future with the advent of the Square Kilometer Array (SKA), populating the GeV sky.
\end{abstract}



\section{Introduction}

The $\gamma$-ray sky in the GeV energy range is dominated by active galactic nuclei (AGNs), as revealed in the past years by the \emph{Fermi}/Large Area Telescope (LAT). Most of  the AGNs detected in $\gamma$-rays are blazars, namely radio-loud AGNs whose jets are oriented close to our line of sight (i.e. $\le$ 10 degrees). In the 4th \emph{Fermi} AGN catalogue (4LAC, \citealt{2020ApJ...892..105A}) 98$\%$ of the detected objects are blazars, while the remainder 2$\%$ are other types of AGNs, collectively named non-blazar sources. These are 69 objects divided into 41 radio galaxies, 9 Narrow Line Seyfert 1 galaxies, 5 Compact Steep Spectrum radio  sources, 2 steep-spectrum radio quasar, 1 Seyfert galaxy and 11 AGNs, whose specific class is still unclear \citep{2020ApJ...892..105A}.  These object types  are of particular interest, since they can provide information on the role of jets in non-blazar objects.

Among \emph{Fermi} non-blazar AGN, radio galaxies constitute the dominant  class and are particularly interesting since they  are considered appealing laboratories to study the high-energy processes at play in AGNs (see e.g. \citealt{2012ApJ...751L...3G,2018MNRAS.476.5535T}). Indeed, their broad-band spectral energy distribution shows signatures of both accretion and jet-related emission. In particular, while the radio and $\gamma$-ray emission is likely due to a jet, the infrared-to-X-ray continuum is generally found to be dominated by the thermal radiation from accreting matter, such as the accretion disc and the hot X-ray-emitting corona (e.g. \citealt{1998MNRAS.299..449W,2007ApJ...659..235G,2011ApJ...740...29K}). Furthermore, they provide a  test bed  where  jet progressive  misalignment can be studied and analyzed. Radio galaxies or misaligned blazar can also be found among still unidentified $\gamma$-ray AGN or between objects wrongly classified as blazars. 

For the two nearby and very extended sources Centaurus A and Fornax A (angular size $\gtrsim 1^\circ$), \emph{Fermi}/LAT was able to detect $\gamma$-ray emission associated with the lobes \citep{2010Sci...328..725A,2016ApJ...826....1A}. This allowed, for the first time, to spatially resolve the emission coming from the lobes and the core even in the GeV domain. Interestingly, the core contribution to the total $\gamma$-ray flux was found to be <14\% in Fornax A \citep{2016ApJ...826....1A}. A thorough SED modeling of the lobes emission in the radio and $\gamma$-rays \citep{2015MNRAS.446.3478M,2016ApJ...826....1A} concluded that both leptonic and hadronic models can explain the observational properties, with a combination of the two being the most probable explanation for the case of Fornax A. Nevertheless, this would imply a high energy density for protons, or filaments with a high gas density and magnetic field energy \citep{2015MNRAS.446.3478M}. Recent studies reconsidered the leptonic model, demonstrating how inverse Compton of the radio-emitting electrons off the ambient optical radiation field can be sufficient to account for the GeV emission in a set of four radio galaxies \citep{2019MNRAS.485.2001P,2019MNRAS.490.1489P,2020MNRAS.491.5740P}. 

We present here a newly discovered radio galaxy, namely IGR\,J18249-3243, which belongs to the complete sample of AGN detected by \emph{INTEGRAL}/IBIS in its first years of survey (Malizia et al. 2009). It is associated to  PKS\,1821-327 which has been classified as AGN of unknown type in the 4th \emph{Fermi} catalogue (4FGL, \citealt{2020ApJS..247...33A}). In this work, we discuss its properties, and how the GeV gamma ray emission might be mostly produced in the lobes. In addition, we present a list of new radio galaxies showing GeV emission in the 4FGL catalogue, whose classification was possible thanks to images from the latest radio surveys. 

In this work we adopt the standard cosmological parameters $H_0=71$ km/s/Mpc, $\Omega=0.27$, $\Omega_\Lambda=0.73$, analogously to our previous works \citep{2016MNRAS.461.3165B}, and the convention $S\propto\nu^\alpha$ for the spectral index definition.


\begin{figure*}
\includegraphics[width=\textwidth]{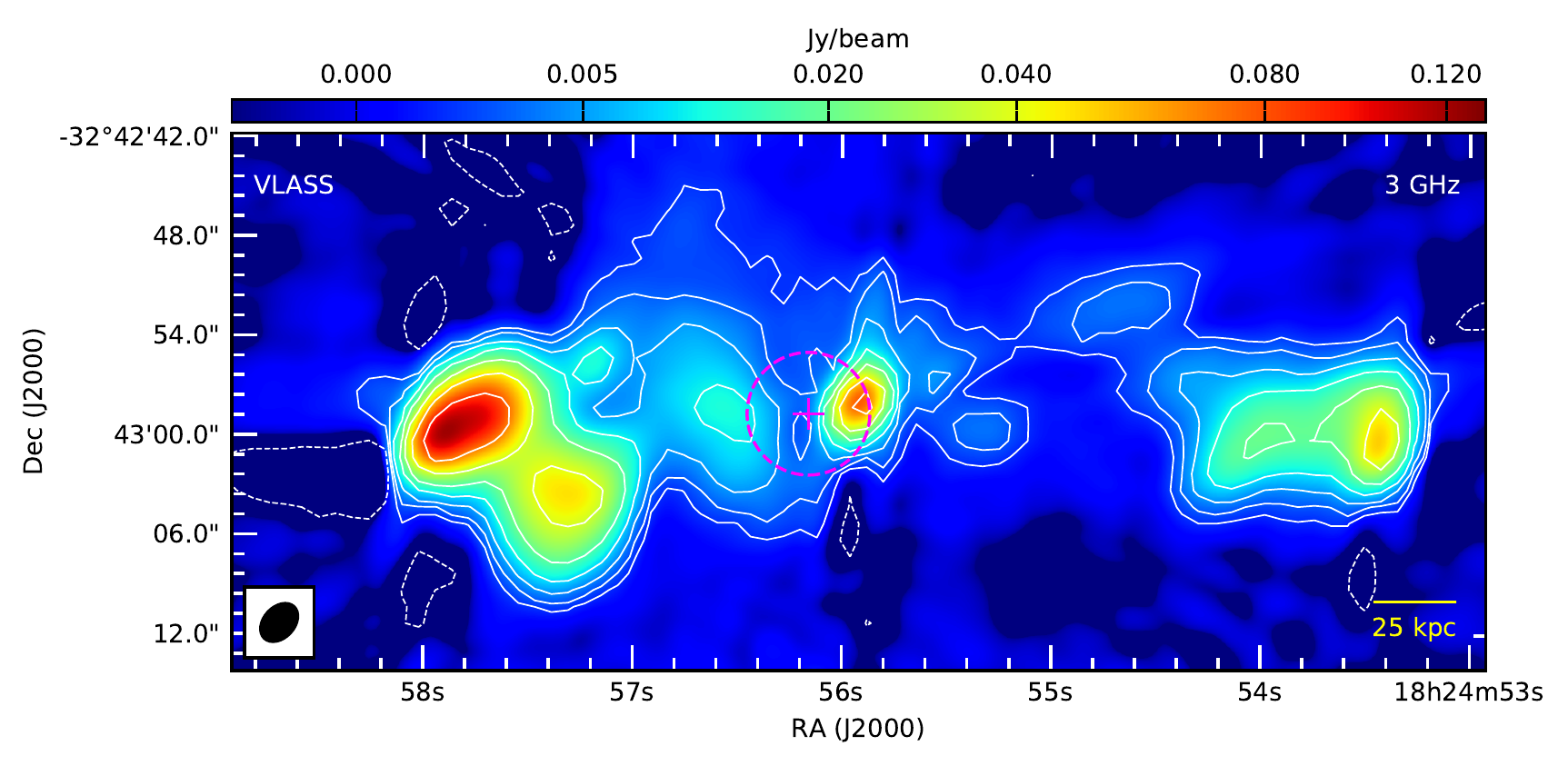}
 \caption{VLASS image of IGR\,J18249-3243. The cross and the dashed circle indicate the position of the X-ray core as measured by \emph{Swift}/XRT, with the corresponding uncertainty. The contours are multiple of the image RMS, namely 3$\times$RMS$\times$(-1, 1, 2, 4, 8, 16, 32, 64). The RMS is 0.411 mJy/beam. The beam is reported in the lower-left corner.}
\label{fig:VLASS}
\end{figure*}


\begin{figure}
\includegraphics[width=\columnwidth]{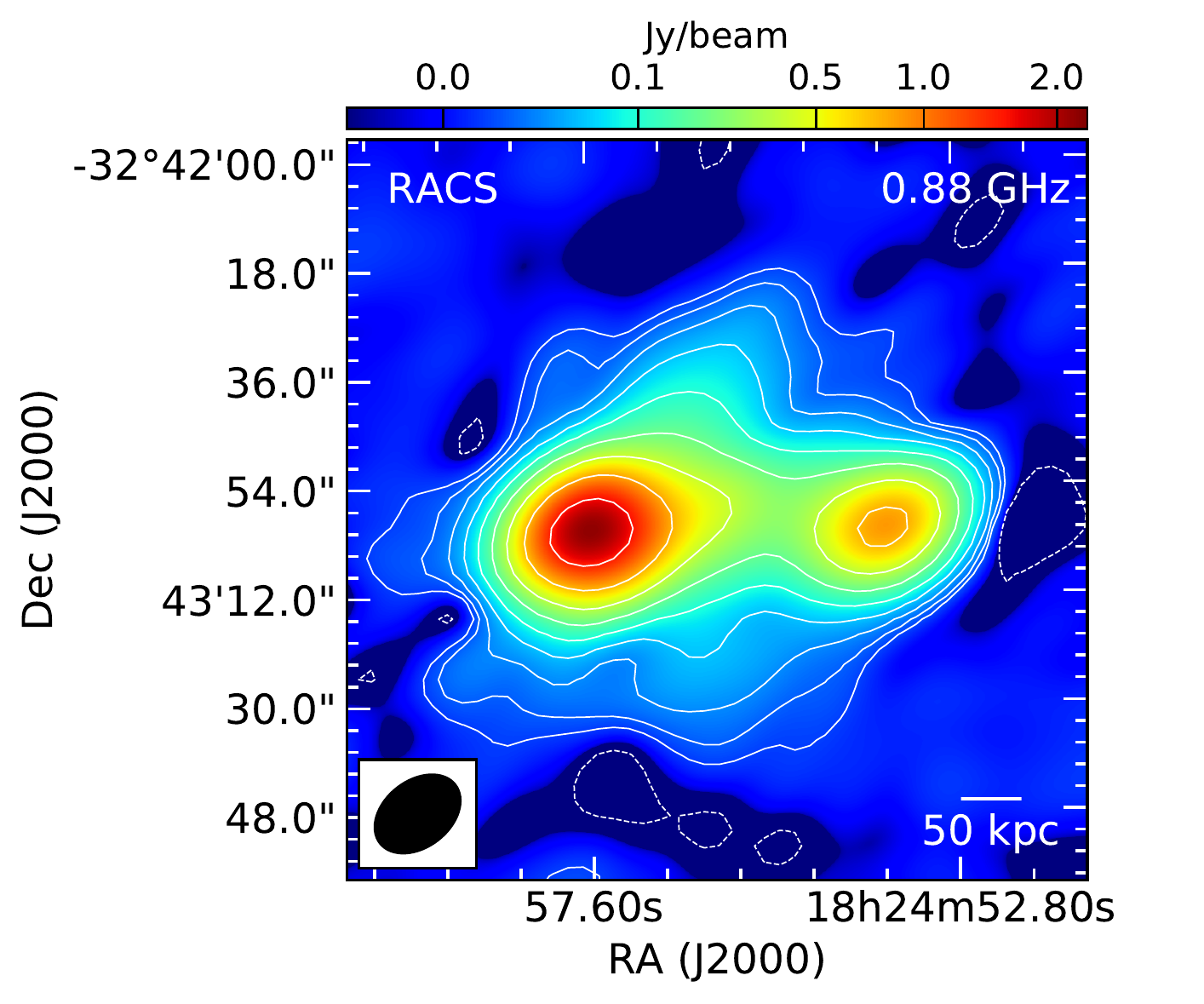}
 \caption{RACS image of IGR\,J18249-3243 at 0.88 GHz. The contours are multiple of the image RMS, namely 3$\times$RMS$\times$(-1, 1, 2, 4, 8, 16, 32, 64, 128, 256). The RMS is 2 mJy/beam. The beam is reported in the lower-left corner.}
\label{fig:RACS}
\end{figure}


\section{The radio galaxy IGR\,J18249-3243}

IGR J18249-3243 was first detected by \emph{INTEGRAL}/IBIS and reported as a new AGN by Bassani et al. (2006) who tentatively associated it to the poorly studied radio source PKS 1821-327. The X-ray counterpart measured by \emph{Swift}/XRT, needed to confirm this radio association and to provide an optical classification throughout follow up observations, was reported by \cite{2007ATel.1273....1L}. The X-ray core is located at RA(J2000)= 18:24:56.11, Dec(J2000)=  --32:42:58.9 with a positional uncertainty of 3.7 arcsec and thus is around 15-20 arsec away from the radio source position (depending on the radio catalogue used). 
In due course, \cite{2009A&A...495..121M} observed the optical counterpart of IGR J18249-3243 with the 1.5 m at the Cerro Tololo Interamerican Observatory (CTIO) in Chile, providing for the first time the source redshift (z=0.355), optical spectrum and AGN class (Seyfert 1, showing broad emission lines).
At the same time \cite{2009A&A...493..893L} explored the broad band characteristics of the system (from radio to X-rays) finding that the source is radio loud, has a complex radio morphology and a steep radio spectrum, while confirming at the same time the AGN nature of the system. By using the combined {\emph{XMM}}/pn--{\emph{INTEGRAL}}/IBIS spectrum, \cite{2014ApJ...782L..25M} further explored the high energy properties of the source finding that a simple power law ($\Gamma$=2), with no evidence for a high energy cut-off, well describes the source spectrum from 0.2 to 200 keV. 

So far, IGR\,J18249-3243 has never been reported as a \emph{Swift}/BAT source, probably due to its location on the galactic plane where the \emph{INTEGRAL}/IBIS coverage is much deeper. 
The most recent update on this source is its detection as a gamma ray emitter, since it is now listed in 4LAC as a non-blazar object \citep{2020ApJ...892..105A}. The GeV $\gamma$-ray spectrum can be described by a simple power law with photon index 2.227$\pm$0.122 and a 0.1-100 GeV flux of 3.8$\pm$0.81 $\times$ 10$^{-12}$ erg cm$^{-2}$ s$^{-1}$. The $\gamma$-ray photon index and luminosity ($1.6\times 10^{45}$ erg s$^{-1}$) locate the galaxy in a region of the ($L_{\gamma},\Gamma$) diagram mostly populated by blazars (see Fig. 11 in \citealt{2020ApJ...892..105A}). Since the optical spectrum of the source shows emission lines, the most likely association would be with a flat spectrum radio quasar; however, the steep radio spectral index, as estimated from surveys (see next section), is not compatible with this classification. Before this work, no radio image  has been able to fully resolve its  structure nor the source has been studied in some details to allow a more specific characterisation of its radio components.


\begin{table*}
\caption{Radio flux densities for the different components of IGR\,J18249--3243. A corresponds to the Eastern lobe, C to the core, and B to the Western lobe.}
\centering
\begin{tabular}{cccccccccccc}
\hline
Telescope   & Frequency     & Resolution        & RMS           & Component     & Flux density      \\    
            & (GHz)         & (arcsec)          & (mJy/beam)    &               & (mJy)             \\
\hline
VLASS       & 3             & 2.6$\times$1.8    & 0.37          & A             &  942$\pm$94      \\
            &               &                   &               & B             &  333$\pm$33      \\
            &               &                   &               & C             &  118$\pm$12      \\
ATCA        & 5.5           & 4.88$\times$1.17  & 0.21          & A             &  603$\pm$30     \\
            &               &                   &               & B             &  198$\pm$10     \\
            &               &                   &               & C             &  122$\pm$6      \\
ATCA        & 9             & 4.88$\times$1.17  & 0.47          & A             &   298$\pm$15     \\
            &               &                   &               & B             &  90.3$\pm$4.5    \\
            &               &                   &               & C             &   111$\pm$6      \\
ATCA        & 17            & 4.88$\times$1.17  & 0.57          & A             &  32.4$\pm$1.6    \\
            &               &                   &               & B             &   6.7$\pm$0.3    \\
            &               &                   &               & C             &  78.4$\pm$4.1    \\
\hline
\end{tabular}
\label{observations}
\end{table*}

\begin{figure}
\includegraphics[width=\columnwidth]{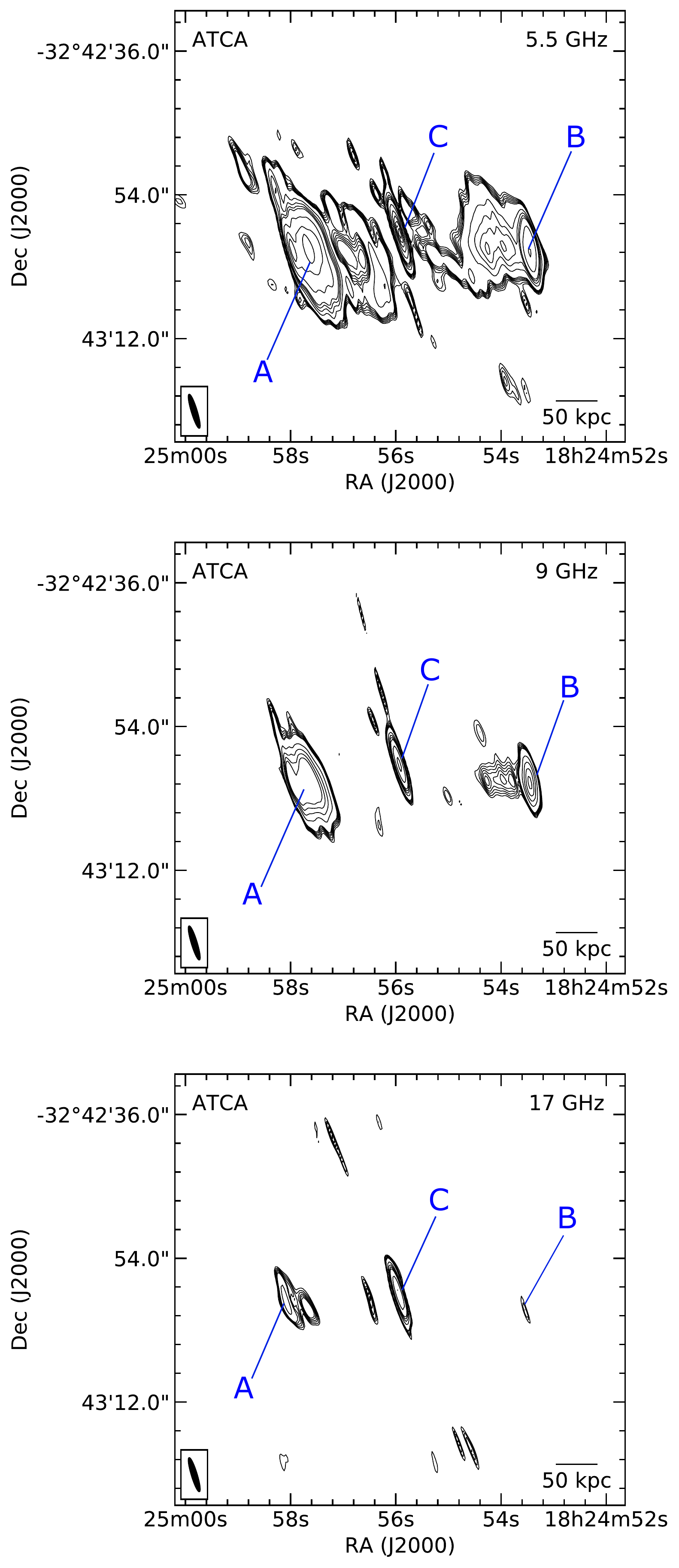}
 \caption{ATCA images of IGR\,J18249-3243 at 5, 9, and 17 GHz. The contours are multiple of the image RMS, namely RMS$\times$(5, 6, 7, 8, 9, 10, 20, 30, 40, 50, 60, 70, 80, 90, 100, 200, 300, 400, 500). The RMS is 0.21, 0.47, and 0.37 mJy/beam for the three frequencies, respectively. The beam is reported in the lower-left corner.}
\label{fig:ATCA}
\end{figure}


\begin{figure}
\includegraphics[width=\columnwidth]{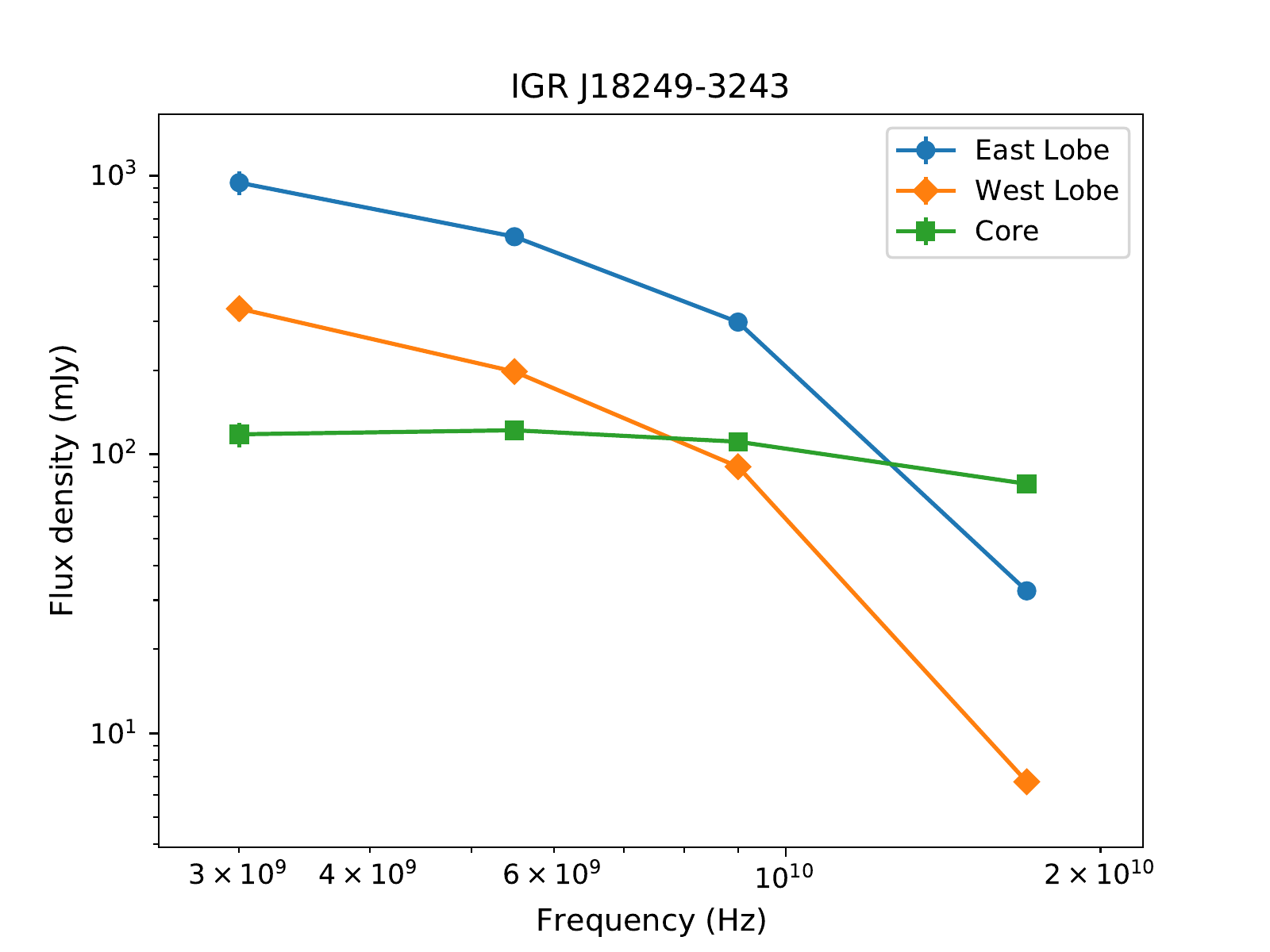}
 \caption{Radio SED for the lobes and core of IGR\,J18249--3243, built with data from VLASS (3 GHz), and ATCA data from our campaign (5.5, 9, and 17 GHz).}
\label{fig:radioSED}
\end{figure}


\subsection{Radio characterisation}

The source was in the footprint of several surveys in the radio band, among which the NRAO VLA sky Survey (NVSS, Condon et al. 2008, 1.4 GHz), the Parkes-MIT-NRAO survey (PMN, \citealt{1993AJ....105.1666G}, 4.8 GHz), the Galactic and extra-galactic all-sky MWA survey (GLEAM, \citealt{2015PASA...32...25W}, 74-231 MHz), and in the TIFR GMRT Sky Survey (TGSS, \citealt{2017A&A...598A..78I}, 150 MHz). However, none of these were able to fully resolve the source morphology. The NVSS image, at a resolution of 45\arcsec, shows a single component with an elongation towards West. Its total flux density is 4080$\pm$204 mJy, resulting in a radio power of $1.26\times10^{27}$ W/Hz. The TGSS, at a slightly higher resolution of 41\arcsec$\times$25\arcsec, better resolves the elongated structure showing an enhanced flux density on the Eastern side. A projected linear size of about 2 arcmin can be estimated. The overall spectral index, considering the flux densities from NVSS and PMN, is steep (--0.77).

Lately, with the advent of new generation radio telescopes, more sensitive and accurate surveys are scanning the sky: the VLA Sky Survey (VLASS, Lacy et al. 2020), and the the Rapid ASKAP Continuum Survey (RACS). The angular resolution is $\sim$2.5\arcsec\, for VLASS, while $\sim$15\arcsec\, for RACS. The sensitivity is at a similar sub-mJy level of $\sim$0.12 mJy/beam for the first epoch of VLASS, and $\sim$0.25 mJy/beam for RACS. However, thanks to the lower frequency and shorter minimum baseline, RACS can detect larger structures than VLASS: the largest angular scale visible by VLASS is $\sim$1\arcmin, versus $\sim$1\degr\, for RACS. Together, they provide a full coverage of the sky. We browsed the first epoch of VLASS quick look images\footnote{\hyperref[https://science.nrao.edu/vlass/data-access/vlass-epoch-1-quick-look-users-guide]{https://science.nrao.edu/vlass/data-access/vlass-epoch-1-quick-look-users-guide}} and found that the angular resolution is sufficient to disclose the morphology of IGR\,J18249-3243. In figure \ref{fig:VLASS}, the source is resolved into three radio components, which all together present the typical morphology of a radio galaxy, with a central core and two lobes. The central component is located within the positional uncertainty of the X-ray coordinates from \emph{Swift}/XRT (pink, dashed circle in Fig. \ref{fig:VLASS}), thus consistent with the core. The total angular scale of the source, as measured between lobes extremities, is 64\arcsec, corresponding to 317 kpc at the source redshift. The surface brightness, larger for the lobes and slightly dimmer for the core, is suggestive of the FRII class. Indeed, the core flux density is 118$\pm$12 mJy, while the Eastern (A) and Western (B) lobes show values of 942$\pm$94 mJy and 333$\pm$33 mJy, respectively. This results in a core dominance ($S_{core}/(S_{tot}-S_{core})$) of 0.09. This value is among the lowest ones found for FRI or FRII radio galaxies previously detected by \emph{Fermi}/LAT \citep{2010ApJ...720..912A}, suggesting a larger viewing angle. The RACS image at 0.8 GHz has a lower resolution (16\arcsec$\times$11\arcsec), nevertheless it resolves the source in two extended components, corresponding with the two lobes (see Fig. \ref{fig:RACS}). At the core position, as derived from the VLASS images, the emission is partly blended with the Eastern lobe. The better capability of the RACS to recover extended emission allows to detect two symmetric tails of emission, on the Northern and one on the Southern side, suggestive of deflected plasma backflow from the lobes (see e.g. \citealt{2020MNRAS.495.1271C}). 

\subsubsection{ATCA observations}
\label{sec:ATCA}

Given the resolved morphology revealed by VLASS, we targeted the source with ATCA at 5.5, 9, and 17 GHz. We performed a 5-hour ATCA observing run as part of the the project C3412 (PI: Ricci) at 5.5, 9 and 17 GHz. The array configuration was 6A, providing an angular resoluton of 4.88$\times$1.17 arcsec at 5.5 GHz, 2.93$\times$0.83 at 9 GHz, and 3.02$\times$0.42 at 17 GHz. The raw RPFITS files were read into {\tt MIRIAD}, flagged for RFIs and calibrated following the same scheme used for AT20G data. The bandpass calibrator was 0537$-$441, the primary calibrator 1934$-$638, and phase calibrator 1759$-$39. As the phases were still very unstable after the standard phase calibration because of atmospheric turbulence, three rounds of phase self-calibration were performed inside {\tt{MIRIAD}} to improve phase stability and thus the restored image quality. Finally, images at 9 and 17 GHz were tapered and restored applying the same beam for the 5.5 GHz one, in order to restore the resolved emission from the lobes. The obtained images are shown in Fig. \ref{fig:ATCA}. 

These observations, at a resolution and sensitivity similar to VLASS, could confirm the previously discussed morphology, and provide the frequency coverage necessary to study the spectral properties of the different components. 
Flux densities for the three components, together with information about the images collected from surveys, are presented in table \ref{observations}. We built a spectral energy distribution for the different components (see Fig. \ref{fig:radioSED}). We found spectral indices between 5.5 and 9 GHz as expected, i.e. steep for the lobes (--1.43$\pm$0.14 for A, and --1.6$\pm$0.14 for B) while flat for the core (--0.19$\pm$0.15), considering a value of --0.5 as boundary between steep and flat spectra. Overall, the discussed radio and optical properties suggest that this source is seen at an intermediate angle between the plane of the sky and the jet axis, allowing to detect broad emission lines in the optical band, but also a symmetric morphology with VLASS and ATCA, typical of radio galaxies.


\subsection{Broad-band SED and the origin of the $\gamma$-ray emission}

We performed an analysis of the broad-band SED, in order to understand which emission mechanisms dominate the different energy ranges. In addition to the radio data described in the previous section, we made use of \emph{XMM}/pn data to cover the X-rays energy range, and the previously mentioned \emph{INTEGRAL}/IBIS and \emph{Fermi}/LAT survey data for the hard X-rays and $\gamma$-ray ones, respectively. The fluxes and upper limits from \emph{Fermi}/LAT were cross-checked by inquiring the online version of the 4FGL catalogue on VizieR\footnote{\hyperref[https://vizier.cds.unistra.fr/viz-bin/VizieR]{https://vizier.cds.unistra.fr/viz-bin/VizieR}}. The source was not reported as variable in the latter. 
In addition, we browsed archival data with the ASI Science Data Center (ASDC) online tool\footnote{\hyperref[http://tools.asdc.asi.it/]{http://tools.asdc.asi.it/}} to cover the IR/Optical bands. However, photometry results to be contaminated by an intervening star, thus could not be used for our target SED. The collected fluxes are reported in Tab. \ref{tab:high}.


\begin{table}
    \centering
    \begin{tabular}{lcc}
    \hline
    Instrument              &   Band        &   Flux        \\
                            &               &   ($\times 10^{-13}$ erg/cm$^2$/s) \\
    \hline
    \emph{XMM}/pn           & 0.2-0.5 (keV) & 1.55$\pm$0.14 \\
                            & 0.5-1         & 6.53$\pm$0.60 \\
                            & 1-2           & 13.0$\pm$1.2  \\
                            & 2-4.5         & 22.1$\pm$2.2  \\
                            & 4.5-12        & 37.2$\pm$3.9  \\
    \emph{INTEGRAL}/IBIS    & 20-40 (keV)   & 39 --1/+4         \\    
                            & 40-100        & $<$42       \\    
    \emph{Fermi}/LAT        & 0.05-0.1 (GeV)  & $<$58.2             \\ 
                            & 0.1-0.3       & $<$10.9               \\ 
                            & 0.3-1         & 6.92 --3.44/+3.41     \\
                            & 1-3           & 6.95 --1.90/+1.94     \\  
                            & 3-10          & 5.81 --1.73/+1.86     \\
                            & 10-30         & $<$2.13               \\
                            & 30-300        & $<$3.19               \\
    \hline
    \end{tabular}
    \caption{High-energy data collected for the broad-band SED of IGR\,J18249--3243.}
    \label{tab:high}
\end{table}


\begin{figure}
\includegraphics[width=\columnwidth]{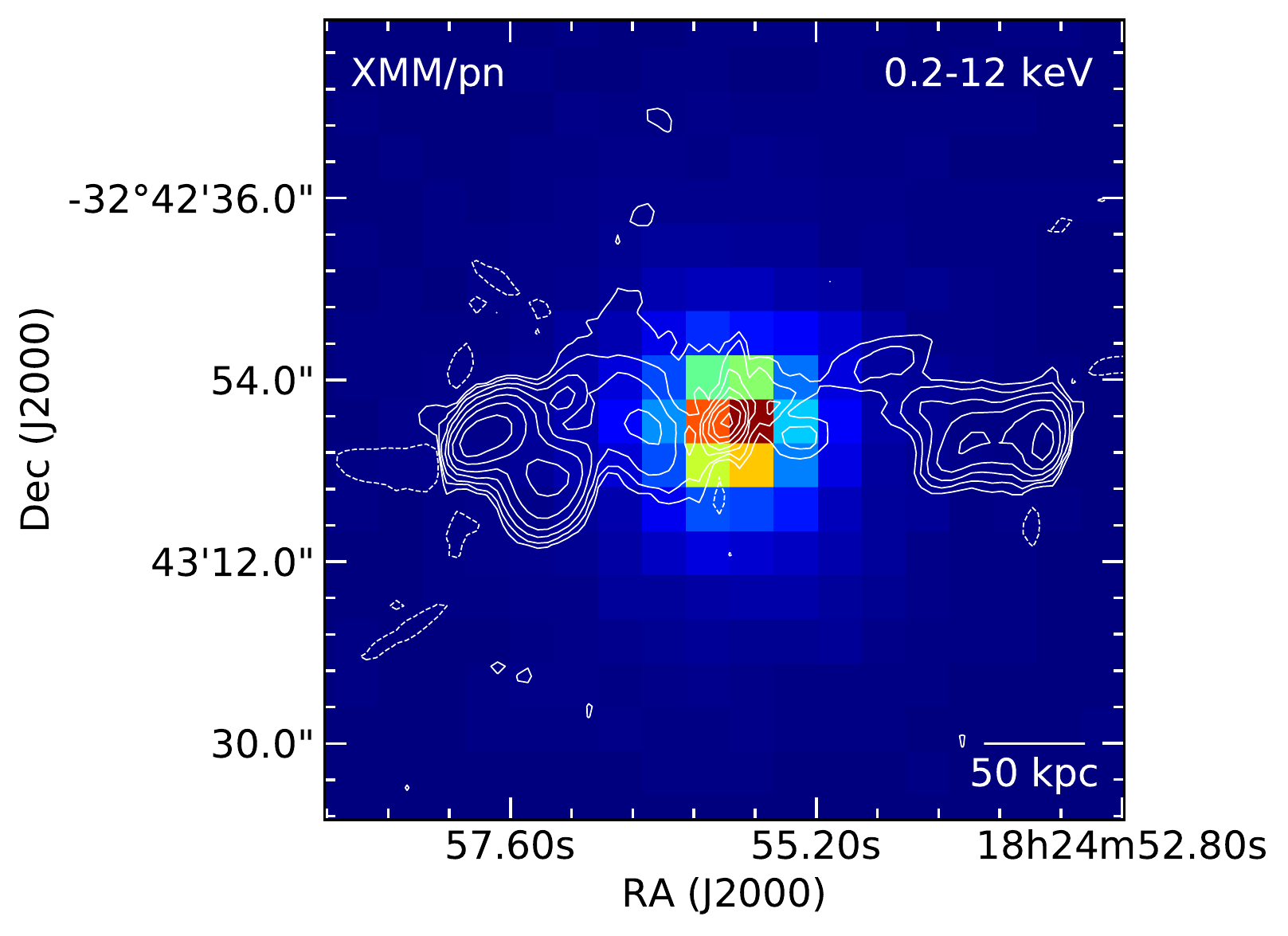}
 \caption{Overlay of the VLASS contours on the \emph{XMM}/pn image (0.2-12 keV). The X-ray emission is dominated by a central point-like source corresponding with the radio core.}
\label{fig:XMM}
\end{figure}


\begin{figure*}
\includegraphics[width=15cm]{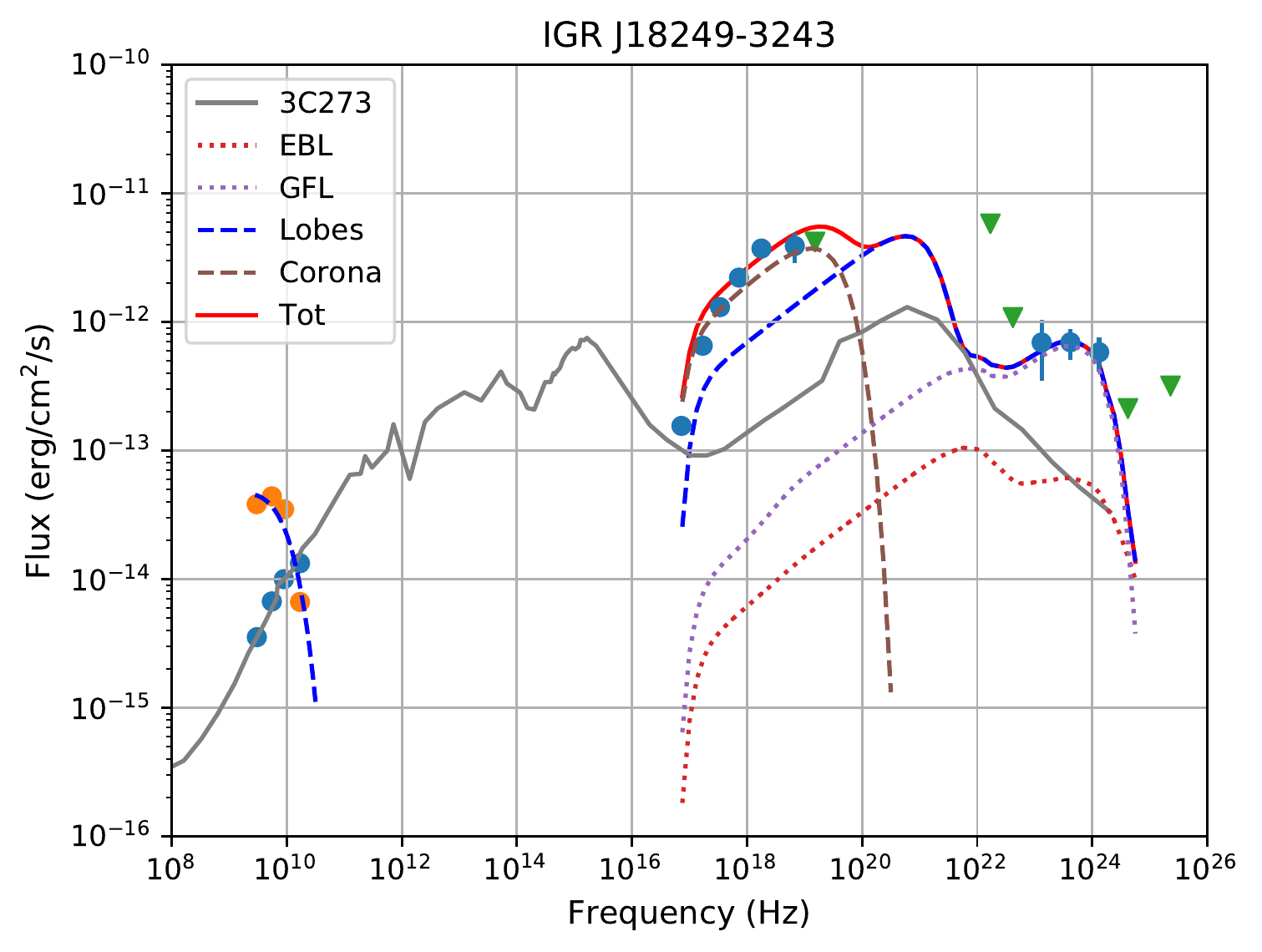}
 \caption{Broad band SED of IGR\,J18249--3243. In the radio frequencies range, orange points indicates the flux densities from the lobes, while blue dots the core ones. The gray solid line indicates the 3C\,273 average spectrum from \protect\cite{2008AA...486..411S}, normalized to the 5 GHz radio core flux. The contribution of the different model components to the SED are indicated with dashed lines of different colors (dotted lines are the lobes sub-components due to foreground and background photon fields), while the red solid line shows the total flux density.}
\label{fig:SED}
\end{figure*}


\subsubsection{Thermal and non-thermal emission from the central AGN}

With these data at hand, we performed a broad-band SED fitting adopting the hybrid model approach by \cite{2011ApJ...740...29K}. This is a phenomenological model for type 1 AGN, considering a thermal component accounting for the IR to hard-X emission produced by the accretion disc, corona, and dusty torus, and a non-thermal component describing the emission from the jet in the radio and $\gamma$-ray band. The purpose of such a simplistic modeling is limited to the quantification of the nuclear contribution to the overall SED, and in particular to verify whether the inner part of the jet can be at the origin of the $\gamma$-ray emission detected by \emph{Fermi}/LAT. Templates available in the literature were considered for both components, namely the average SED for Seyfert 1 built by \cite{2010MNRAS.402..724P} for the thermal component, and the average 3C\,273 SED from \cite{2008AA...486..411S} for the non-thermal one. Analogously to \cite{2011ApJ...740...29K}, we considered the 5 GHz radio fluxes from the core region only as a constraint for the 3C\,273 template, since the blazar-like, non-thermal emission described by the 3C\,273 template is dominated by the core. As for the Seyfert 1 template, the model should have been matched to the infrared-to-X-rays continuum. However, only the X-rays/hard-X rays ranges were covered by meaningful measurements, while the optical and NIR data points were not usable due to the previously mentioned contamination by an intervening star. Thus, we preferred not to use the thermal model for further analysis. The result is shown in Fig. \ref{fig:SED}. 

The non-thermal component is found to be in good agreement with the ATCA and VLASS measurements. However, the three \emph{Fermi}/LAT detections have a flux excess of about an order of magnitude with respect to the values expected from the template. This supposes an additional emission mechanism or region from what found by \cite{2011ApJ...740...29K} for X-ray bright broad-line radio galaxies, where the GeV emission was found to be dominated by the beamed radiation from the relativistic jet. Indeed, in their work the \emph{Fermi}/LAT fluxes were in good agreement with the ones expected from the non-thermal component. We note that the source has a hard-X rays luminosity (14-195 keV) among the highest ones found for the combined \emph{Swift}/BAT+\emph{INTEGRAL}/IBIS radio galaxies sample \citep{2016MNRAS.461.3165B}. 
This possibly indicates the need for 
an additional emitting component in the high-energy portion of the SED. However, because of the lack of reliable measurements in the NIR-to-optical range, the expected yield 
in the soft and hard X-ray region cannot be predicted, and thus the presence of X-ray excess 
(with respect to the expected fluxes from the Sy1 template) cannot be determined.

Finally, we note that previous studies on the radio/$\gamma$-ray connection in the \emph{Fermi}/LAT AGN sample reported a considerable scatter of the GeV emission values for a given radio luminosity \citep{2011ApJ...741...30A}. That result, while establishing a significant correlation between the two quantities, allows for a wide range of luminosities to be expected for $\gamma$-ray counterparts of radio sources, thus mitigating the apparent $\gamma$-ray excess highlighted by the modeling approach proposed by \cite{2011ApJ...740...29K}, and adopted in this work. 


\subsubsection{Lobe $\gamma$-ray emission}
As discussed before, 
analysis of the SED indicates the need for an additional component of $\gamma$-ray emission, 
beyond that of the core, to account for the flux measured by \emph{Fermi}/LAT. In accord with previous studies \cite{2019MNRAS.485.2001P}, an appreciable contribution to the $\gamma$-ray emission could come from the radio lobes. 
Resolved radio maps and spectral X-and $\gamma$-ray measurements of the lobes of several radio galaxies have provided a basis for determining the emission processes, the spectrum of the energetic particles (`cosmic rays'), and the mean value of the magnetic field in the lobes. It has been shown that for a proper spectral analysis the superposed radiation fields in the lobes need to be accounted for (Fornax\,A: \citealt{2019MNRAS.485.2001P}; Centaurus\,A, Centaurus\,B, and NGC\,6251: \citealt{2019MNRAS.490.1489P}). These SED analyses have confirmed the leptonic lobe origin of the $\gamma$-ray emission measured by {\it Fermi}/LAT, thereby constraining the level of hadronic contribution to less than 
$\sim 10-20\%$. 

Radio maps of IGR\,J18249-3243 show the presence of nuclear emission and two asymmetric but roughly comparable, off-center emission regions whose centers are located at $\sim 120$ kpc east and west of the central 
galaxy at a luminosity distance of $1.88$ Gpc. Whereas most of the radio emission is in the lobes, 
X-ray emission is dominated by a central point-like source (see Fig. \ref{fig:XMM}). However, diffuse emission  beyond 
the {\emph{XMM}} point spread function (PSF) cannot be excluded, although not visible in the {\emph{XMM}} 
image. Future {\emph{Chandra}} high spatial resolution images may allow us to provide an estimate on the flux in the lobes region. Neither the {\it INTEGRAL}/IBIS nor the {\it  Fermi}/LAT emission is resolved.

The radio emitting (cosmic ray) electrons (CRe) in the lobes and the central galaxy (jets, disc, 
and halo) emit also X/$\gamma$-ray by Compton scattering off the CMB and the local radiation 
field. Here we model the diffuse non-thermal emission of IGR\,J18249-3243 following \cite{2019MNRAS.485.2001P,2019MNRAS.490.1489P,2020MNRAS.491.5740P}. A precise determination of the ambient photon fields in the lobes is 
needed to predict Compton X/$\gamma$-ray emission by CRe. Radiation fields in the lobes 
include cosmic and local components. Cosmic radiation fields include the CMB and the 
Extragalactic Background Light (EBL). The 
CMB temperature is $T_{\rm CMB} = 2.725\,(1+z)$\,K and energy density $u_{\rm CMB}=0.25\, (1+z)^{4}$\,eV\,cm$^{-3}$. The EBL originates from direct and dust-reprocessed starlight 
integrated over the star formation history of the universe. Adopting a recent updated EBL model, 
based on galaxy counts in several spectral bands \citep{2017A&A...603A..34F}, at $z = 0.355$ 
the EBL photon number density can be numerically approximated as a combination of diluted Planckians, 
\begin{equation}
\lefteqn{
n_{\rm EBL}(\epsilon) ~=~ \sum_{j=1}^8 A_j \,{8 \pi \over h^3c^3} \, \frac{\epsilon^2}{e^{\epsilon/k_B T_j}-1} \hspace{0.5cm}  
{\rm cm^{-3}~ erg^{-1}} }
\label{eq:EBL}
\end{equation}
with: 
$A_1=10^{-5.596}$, $T_1=31.522$\,K; 
$A_2=10^{-7.231}$, $T_2=61.053$\,K; 
$A_3=10^{-9.574}$, $T_3=207.143$\,K; 
$A_4=10^{-11.629}$, $T_4=517.857$\,K;  
$A_5=10^{-13.249}$, $T_5=2088$\,K; 
$A_6=10^{-14.536}$, $T_6=4495$\,K; 
$A_7=10^{-16.496}$, $T_7=10000$\,K;
$A_8=10^{-18.124}$, $T_7=23200$\,K. 


\begin{table*}
    \centering
    \scalebox{0.93}{
    \begin{tabular}{ccccccccccccc}
    \hline
    4FGL ID			 & Other ID				    & z      & Survey/Ref.                   &   LS          & LS        & Total flux density  & Total radio power  & Class  \\
                         &      &        &      &   (arcsec)    & (kpc)   & (mJy)   & log(W/Hz)  &  \\
    \hline
    J0119.6+4158	& NVSS\,J012002+420008	    & 0.109	 & VLASS/LoTSS                  & 110   & 216   & 255.0$\pm$8.8         & 24.83  & FRI    \\
    J0929.3--2414	& NVSS\,J092928--241632		& --	 & VLASS/RACS                   & 130   & --    & 264.9$\pm$8.9         & --     & FRI    \\
    J1344.4--3656	& NVSS\,J134423--365627		& --	 & VLASS/RACS                   & 54    & --    & 508.4$\pm$17.9        & --     & FRII   \\
    J1455.4--3654	& NVSS\,J145509--365519		& 0.095  & VLASS/RACS                   & 210   & 365   & 1070.3$\pm$34.4       & 25.34  & FRI    \\
    \hline  
    J0038.7--0204  & 3C\,17                   & 0.220  & \cite{1999AAS..140..355M}    &  38   & 181   &  6187.8$\pm$239.1	  & 26.85  & FRI/II \\
    J0522.9--3628  & PKS\,0521--36            & 0.055  & \cite{2015MNRAS.450.3975D}   &  50   & 53    &  11882.9$\pm$356.5    & 25.89  & FRI/II \\
    J1236.9--7232  & PKS\,1234--723           & 0.0234 & \cite{2002MNRAS.331..717L}   &  840  & 391   &  1341$\pm$67          & 24.20  & FRI    \\
    J2334.9--2346  & PKS\,2331--240           & 0.0475 & \cite{2017AA...603A.131H}    &  1320 & 1150  &  1141.0$\pm$57.0      & 24.75  & FRII   \\
    \hline
    \end{tabular}
    }
    \caption{New radio galaxies from the latest radio surveys (top 4 objects) or from the literature (bottom 4 objects) with emission in the GeV band detected by \emph{Fermi}/LAT. In the last two columns we report the total flux density and radio power (for sources with 
    known redshift) 
    as measured at 1.4 GHz from NVSS, or at 0.8 GHz from RACS for the Southern source PKS\,1234--723.}
    \label{tab:newRG}
\end{table*}


The local radiation fields (Galaxy Foreground Light, GFL) arise from the IR and optical humps 
of the central galaxy. Its bolometric optical luminosity is $L_{\rm opt} \sim 10^{46}$ erg s$^{-1}$, estimated from $V \simeq 15.86$ mag (\citealt{2019PASA...36...33O}; however, this magnitude may be 
affected 
by a foreground star: \citealt{2009A&A...495..121M}) and applying the bolometric correction BC$_V=-$0.85 \citep{2006MNRAS.368..877B}. The bolometric total IR (8$-$1000 $\mu$m) luminosity, $L_{\rm IR} 
\sim 4 \cdot 10^{45}$ erg s$^{-1}$, is estimated from $l_B$ (monochromatic blue luminosity, 
$l_B = L_B / 6.88E+14\,{\rm Hz}$) 
using 
the $L_{\rm FIR}$--$l_B$ relation by \cite{1998LNP...506..551T} for late-type spirals (in the far-IR 
60$-$100 $\mu$m, band), assuming $L_{\rm IR} \sim 1.7\, L_{\rm FIR}$ \citep{2007A&A...463..481P}. However, 
given that the contribution of the scattered IR field to the LAT $\gamma$-ray data is very small, the actual value of $L_{\rm IR}$ is unimportant. The IR and optical parameters allow us to model the 
GFL; in our calculations we take $T_{\rm gal,\,OPT} = 2900$\,K and $T_{\rm gal,\,IR} = 29$\,K \citep{2019MNRAS.485.2001P}. The lobe X-ray data are spatial averages, so we correspondingly compute volume-averaged Compton/GFL yields based on the fact that lobe sizes and projected distances (from the central galaxy) are much larger than the size of the galaxy, which we treat as a point source \citep{2019MNRAS.490.1489P}. 

Radio emission in the lobes is by electron synchrotron in disordered magnetic fields whose mean 
value $B$ is taken to be spatially uniform, and X-$\gamma$ emission is by Compton scattering 
off the CMB and optical radiation fields. The calculations of the emissivities from these processes 
are standard  (e.g. \citealt{2019MNRAS.485.2001P}). Assuming steady state, the CRe spectrum is assumed 
to be spatially isotropic, truncated--power-law (PL) distribution in the electron Lorentz factor, $N_e(\gamma) = N_{e0}\, \gamma^{-q_e}$ in the interval $[\gamma_{min},\, \gamma_{max}$]. 
The photoelectrically absorbed ($N_H = 1.2 \cdot 10^{21}$ cm$^2$, Landi et al. 2008) X-ray 
spectrum is used to determine $N_{e0}$ and $q_e$; alternatively, the $\gamma$-ray spectrum can 
be used if the GFL field is precisely known and no appreciable $\pi^{0}$-decay emission is assumed (as deduced in other analyses of radio lobe SEDs; \citealt{2019MNRAS.485.2001P,2019MNRAS.490.1489P,2020MNRAS.491.5740P}). Comparison of the predicted and measured radio spectra yields $B$. The radio and $\gamma$-ray high-frequency spectral turnovers reflect the same value of $\gamma_{\rm max}$. 

Before modeling the broad-band SED as outlined above, we must assess the origin of the (the soft and hard) X-ray spectrum in order to identify a Compton/CMB component. 
As discussed above, 
the {\emph{XMM}} map does not show any clear extended emission 
(when account is taken of the detector point spread function) 
and because of insufficient spatial resolution of INTEGRAL/IBIS, we can not resolve hard-X 
emission into core and lobes. Accordingly, X-ray emission is assumed to originate from two 
regions: a central point-like AGN corona (and disc) and two outer extended 
lobes; however emission from the lobe is estimated to be below the {\emph{XMM}} sensitivity. 

Therefore a viable SED model comprises a nuclear component, modeled as a cutoff-PL with a photon index $\Gamma=1.6$ (compatible with values in \citealt{2016MNRAS.458.2454L}), 
a cutoff energy $E_c = 250$ keV, representing the AGN corona, and an extended 
lobe component, modeled as Compton/CMB emission.

We parametrize the relative contribution of the two components in terms of the fraction, $f$, of lobe emission at 1\,keV. 
From the fact that emission from the lobes is not seen 
in the {\emph{XMM}} surface brightness map, we choose a low value of $f$ of a few percent. The model 
overlaid on data in Fig. \ref{fig:SED} has $f = 2.5$\%\footnote{
Emission parameters are: $N_{e0} = 5.9\,E-3$ cm$^{-3}$, $q_e = 2.32$, $\gamma_{\rm min} = 100$, $\gamma_{\rm max} = 6\,E+4$, $B = 0.65\,\mu$G; $V = 15.86$\,mag \citep{2019PASA...36...33O}, 
$(B-V) = 0.4$\,mag 
(typical of spiral discs which Seyfert 1 nuclei resemble).}

The GFL provides the dominant contribution to the Comptonized starlight, whereas the EBL 
component is sub-dominant (similar to Fornax A, \citealt{2019MNRAS.485.2001P}). We note that, 
in principle, the GFL contribution has no degrees of freedom once, for a given CRe spectrum, the galaxy luminosity is known: 
However, in this case 
the latter is poorly known so some freedom exists in matching the {\it Fermi}/LAT data. (A reliable central galaxy luminosity would clearly help in determining the CRe spectral normalization hence the lobe contribution to the X-ray emission.) Given the uncertainty on the host V-mag (and the LAT data, 
as well), a putative minor contribution from the inner jet ($\sim$10\%/5\% at the lowest/highest 
energy LAT point; see previous section) can be easily accommodated in the overall SED 
model. A 
moderate level of uncertainty of just 0.1 mag fainter V-magnitude would be equivalent to this level of contribution. 

Given the 
substantial 
uncertainties stemming from the lack of spatial distribution of the emission, the above SED 
model -- which is not based on exact statistical analysis -- is meant to be merely suggestive 
as to the likely origin of (most of) the observed $\gamma$-ray emission. 


\section{New radio galaxies among \emph{Fermi}/LAT detected AGN}
\label{sec:newRG}

The discovery of yet another radio galaxy among \emph{Fermi} detected AGN, raises the question of how many such objects are still hidden among the general population of GeV emitters, mainly due to the lack of resolution and sensitivity of classical radio surveys  used so far for identification/classification purposes of \emph{Fermi}/LAT sources. Furthermore, the fact that IGR J18249-3243 is an FR II type object queries   the evidence that FRI radio galaxies are preferentially seen among AGN emitting at  high $\gamma$-ray energies \citep{2010ApJ...720..912A}.
Recent radio surveys like VLASS and RACS in the GHz domain, and the LOFAR Two-metre Sky Survey (LoTSS, \citealt{2017A&A...598A.104S}) in the MHz domain, have provided a significant improvement both in sensitivity and resolution over previous ones, offering the opportunity to discover new radio galaxies among \emph{Fermi}/LAT AGN. This also makes possible to reassess the ratio of FR I versus FR II types, and therefore provides more insight into the origin of the $\gamma$-ray emission in radio galaxies.

In the following, we provide observational evidence that a small sample (8 objects) of \emph{Fermi}/LAT AGN can be classified as new radio galaxies either from the literature/available databases or by means of our own analysis using VLASS, RACS, and LoTSS images. We note that the `AGN' classification used in \emph{Fermi}/LAT catalogues is defined as to exclude blazars or candidates of such class, thus already being a good indication of a possible misaligned nature of the source.


\subsection{New GeV-emitting radio galaxies from the latest radio surveys}
\label{sec:newRG}

Recently, \cite{2020JHEAp..27...77C} compiled a list of candidate Gev-emitting misaligned AGN (MAGN) by cross-correlating the fourth \emph{Fermi}/LAT catalog with the NVSS and SUMSS radio surveys. At the resolution and sensitivity of these surveys (HPBW$\approx$45\arcsec, RMS$\approx$0.5-1.2 mJy/beam), they could identify 48 objects with an angular extension larger than the survey beam. To improve their classification, we cross-correlated their list of candidate MAGN with the latest  VLASS and RACS radio surveys, in order to have higher sensitivity and resolution images. Among the objects listed by \cite{2020JHEAp..27...77C} that were mapped by these recent surveys, 4 clearly show a resolved, double-lobed morphology in the images (see Tab. \ref{tab:newRG}). All of them were detected in VLASS, and three out of four also in the RACS survey. 

The collected images are presented in Fig. \ref{MAGN1} and \ref{MAGN2}. For each source, we report the VLASS image with RACS contours superimposed (left panel), and the Pan-STARRS or DSS image (right panel), again with RACS contours overlaid. The latter show the extended lobes emission, not detected by VLASS because of the different survey design. The optical image helps in verifying that the radio core position corresponds with the host galaxy, confirming the double-lobed morphology (and thus the radio galaxy classification) for all of them. Given the declination of 4FGL\,J0119.6+4158, RACS data were not available. We thus made use of the LoTSS DR2 image at 150 MHz, at an angular resolution of 6 arcsec (Shinwell et al. submitted).

The optical redshift is available in the literature for two sources only. For 4FGL\,J1455.4-3654, Simbad and NED report a redshift of $z=$0.095. For 4FGL\,J0119.6+4158, \cite{2020A&A...643A.103P} report a redshift of $z=$0.109 and an optical classification as a BL Lac galaxy-dominated source (bzg); these objects are usually reported as  BL Lac  in the literature, but with an  energy distribution  dominated by emission from the host galaxy. However, as the authors state, not all bzg are expected to be genuine blazar  as some could be moderately bright AGNs, such as radio galaxies, with non-thermal emission without strong relativistic beaming; this turns out to be the case for 4FGL J0119.6+4158/2MASX J01200274+4200139. 

The linear size and radio power are reported in Tab. \ref{tab:newRG}, together with the suggested FR classification based on the radio power at 1.4 GHz and morphology.


 \subsection{New GeV-emitting radio galaxies from the literature}
 
 In a recent work, \cite{2020APh...11602393A} used the sample of radio galaxies presented in 4LAC to quantify how many FR I versus FR II types emit at gamma ray energies. He assumed as a dividing threshold between the two classes of objects a total 1.4 GHz radio power of 10$^{25}$ W/Hz, finding 28 FR I and 12 FR II objects in addition to PKS\,1718-649, the first young radio galaxy detected in $\gamma$-rays \citep{2009AN....330..193G,2016ApJ...821L..31M}. The latter was classified by \cite{2017ApJ...849...34O} as a GigaHertz-Peaked Spectrum (GPS) source, as well as a Compact Symmetric Object (CSO). 
 Some of the sources listed in \cite{2020APh...11602393A} were classified in the original \emph{Fermi}/LAT sample as AGN, but a closer look at their morphology indicates the classical double-lobed structure. We therefore expand this previous analysis to include all objects from the \cite{2020ApJ...892..105A} list that are classified as non-blazar but not included in \cite{2020APh...11602393A}; we searched for each object the literature or available database to confirm if the radio structure was that of a radio galaxy. As a result of this analysis, we have been able to add to the list of radio galaxies compiled by \cite{2020APh...11602393A} the following 4 objects with their relative classification:

\begin{itemize}

\item 4FGL J0038.7-0204: this radio galaxy, fully described in \cite{1999AAS..140..355M}, displays a peculiar radio morphology: on the south-east side of the nucleus a very bent jet is observed while on the western side a lobe structure is visible with a ring-like shape. The 1.4 GHz radio power of 7.9$\times10^{26}$ W/Hz is at least two orders of magnitude above the division between FR I and FR II types assumed by Angioni (2019), pointing to a FRII classification, but the ambiguous radio morphology led \cite{2009ApJ...695..755M} to define  the  source as another example of the "hybrid” (FRI/FRII) radio galaxy type. The classification of this source was later changed to radio galaxy in the 4LAC update \citep{2020arXiv201008406L}.


\item 4FGL J0522.9-3628: as already reported by \cite{2020ApJ...892..105A}, \cite{2015MNRAS.450.3975D} presented a multi waveband study of this object from radio to $\gamma$-rays -- indeed, it was already present in the 3FGL catalogue by \citealt{2015ApJS..218...23A} -- highlighting its intermediate nature between broad-line radio galaxies (BLRG) and steep spectrum radio quasars (SSRQ). Its power at 1.4 GHz is 7.8$\times$10$^{25}$ W/Hz, slightly larger than the threshold for the FRII class. Its morphology is not completely resolved at the resolution of the available observations, and the source is not yet included in the VLASS survey. \cite{2015MNRAS.450.3975D} identified three components: the core, a jet-like structure towards NW, and a further component towards SE. Their broad-band SED fitting suggests a viewing angle between 6-15 degrees. 


\item 4FGL J1236.9-7232: this source is fully discussed by \cite{2002MNRAS.331..717L} and classified as a FR I on the basis of ATCA images. This is also confirmed by its radio power at 1.4 GHz of 1.58$\times$10$^{24}$ W/Hz, calculated from the flux density reported in the NVSS catalogue. Finally, its noticeable angular size of 14 arcmin translates into a projected linear size of 391 kpc. Also, the source classification is now radio galaxy in the 4LAC update \citep{2020arXiv201008406L}. 


\item 4FGL J2334.9-23469: as pointed out by \cite{2020ApJ...892..105A}, this sources was fully analysed by \cite{2017AA...603A.131H,2018MNRAS.478.4634H} and found to host a blazar-like nucleus in a giant size radio galaxy (1.2 Mpc) displaying a FRII morphology; to reconcile  these two opposite observational evidences, the authors suggest that the source suffered a dramatic change in jet orientation as a consequence of restarting activity. The total power at 1.4 GHz is 5.62$\times$10$^{24}$ W/Hz, slightly below the limit for FRI/FRII classification. Given the peculiar nature of this source, and the suspected reactivation and reorientation of the core, the emission detected by \emph{Fermi}/LAT could be due to the blazar corresponding with its core, that would then pertain to the blazar class rather than to the radio galaxies one.

\end{itemize}


\begin{table}
    \centering
    \scalebox{1.0}{
    \begin{tabular}{ccccccccccccc}
    \hline
    4FGL ID &  CD   & log L$_\gamma$ [0.1-10 GeV] & $\Gamma$    \\
            &       & (erg/s)                     &             \\
    \hline
    J0119.6+4158  &  0.29 &   43.58  &  2.29$\pm$0.16  \\               
    J0929.3–2414  &  0.89 &   -      &  2.22$\pm$0.16  \\
    J1344.4–3656  &  0.24 &   -      &  2.30$\pm$0.11  \\
    J1455.4–3654  &  1.45 &   43.76  &  2.44$\pm$0.14  \\  
    J1824.7-3243  &  0.09 &   44.79  &  2.23$\pm$0.12  \\
    \hline
    J0038.7–0204  &  0.33 &   44.53  &  2.80$\pm$0.11  \\  
    J0522.9–3628  &  0.79 &   44.50  &  2.45$\pm$0.01  \\
    J1236.9--7232 &  1.49 &   42.24  &  2.36$\pm$0.12  \\
    J2334.9--2346 &  0.45 &   43.57  &  2.65$\pm$0.14 \\                
    \hline
    \end{tabular}
    }
    \caption{Collected quantities for the new {\emph{Fermi}}/LAT radio galaxies from VLASS (top half) or the literature (bottom half), used for the comparison with \protect\cite{2010ApJ...720..912A}.}
    \label{tab:CD}
\end{table}


\begin{figure}
\includegraphics[width=\columnwidth]{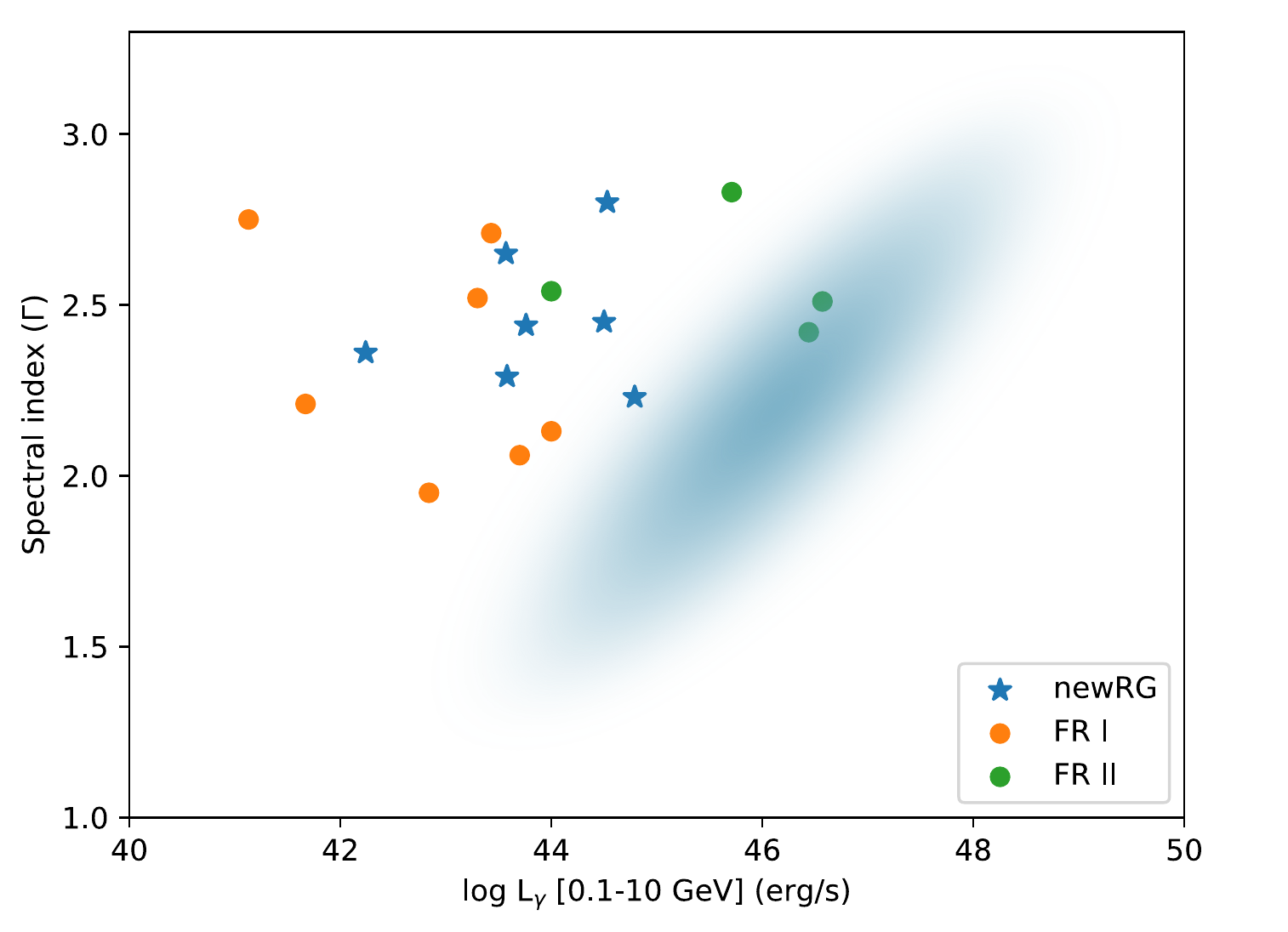}
 \caption{Spectral index at GeV energies ($\Gamma$) as a function of the 0.1-10 GeV luminosity. The locus of blazars is indicated as a shaded area. FR I and FR II from \protect\cite{2010ApJ...720..912A} are in orange and green points respectively, while the newly discovered radio galaxies with a redshift estimate are represented as blue stars.}
\label{fig:Abdo}
\end{figure}


\subsection{Main properties}

We used data from the latest radio surveys and information in the literature to identify further radio galaxies among \emph{Fermi}/LAT AGN. In total we found 9 such objects, including IGR J18249--3243 which has been the main focus of this paper: four can be confidently classified as FR I while 2 as FR II and other two as intermediate FRI/FRII objects.
This allows to update the relative abundance of  FR I and FR II  among GeV emitting radio galaxies. Expanding on the list  of \cite{2020APh...11602393A}, we find a FRI versus FR II  ratio of roughly 2.5 (32 FR I versus  14 FR II plus   two intermediate sources), thus confirming the prevalence of FR I radio galaxies initially found by \cite{2010ApJ...720..912A}, and not yet fully explained in the literature. 

 Among the five newly discovered radio galaxies found in VLASS (IGR\,J18249-3243 plus bottom half of Tab. \ref{tab:newRG}), three have an estimate of the redshift value, making possible the calculation of the $\gamma$-ray luminosity from 4FGL flux values. In addition to these ones, we consider also the four objects found in the literature.  We then used the $\Gamma$ vs log L$_\gamma$ [0.1-10 GeV] (erg/s) diagram (see figure \ref{fig:Abdo}) to display our 7 sources in comparison with the 11 MAGNs  detected by \emph{Fermi}/LAT 
 in the first 15 months of observations and discussed  by \cite{2010ApJ...720..912A}; also shown in the figure is the locus of blazar like objects. The sample of 11 objects included seven FRI and four FRII radio galaxies. The latter showed $\gamma$-ray luminosities similar to blazars, while the FRI ones were typically three orders of magnitude lower.
 It is evident from the figure that our radio galaxies display lower gamma ray luminosities than blazar like objects; in addition we note that our objects have  intermediate values compared to previously known radio galaxies.
 
 \cite{2010ApJ...720..912A}  also estimated the core dominance of MAGNs, proxy of their orientation, finding high values when compared with the 3CRR sample. This indicates a preferred orientation along the jet axis for \emph{Fermi}/LAT MAGNs. Regarding our objects, we used VLASS images to estimate the core dominance at 3 GHz for the five objects discovered from VLASS with this work (top half of Tab. \ref{tab:CD}), while we compiled values from the literature for the others (bottom half of Tab. \ref{tab:CD}). The collected values range from $\sim$0.1 to $\sim$1.5.
 Overall, we conclude that the new GeV emitting radio galaxies found in this work have core dominance values within the range reported by  \cite{2010ApJ...720..912A}, suggesting the same preferred polar orientation for MAGNs.


\section{Final remarks}

\subsection{The emerging population of GeV-emitting radio galaxies}

In this work, we presented evidences of how IGR\,J18249--3243, reported as a new detection in the \emph{Fermi}/LAT survey of \cite{2020ApJ...892..105A} and generically classified as an AGN, is actually a radio galaxy of the FR II type. Its double-lobed structure was missed until now due to the lack of resolution in classical radio surveys such as NVSS and SUMSS. The discovery that IGR 18249-3243 is a radio galaxy of the FRII type raises further considerations, in terms of the incidence of $\gamma$-ray emission on the general AGN population, and more specifically in radio galaxies. Considering our broad-band SED modeling of IGR\,J18249--3243, and its possible GeV emission from the lobes, we can foresee that more FR II objects with a GeV counterpart will emerge from new, deeper, and sharper radio surveys, softening the prevalence of FR I found so far (\citealt{2010ApJ...720..912A}). A first example of the potentiality of these surveys has been provided in Sec. \ref{sec:newRG}, where we could identify 4 new radio galaxies among the 47 candidates ($\sim$10\%) provided by \cite{2020JHEAp..27...77C}.   

As new censuses of $\gamma$-ray sources are built, an emerging population of MAGNs and non-blazar sources is being found \citep{2021arXivF}. In particular, the recently introduced class of FR0 sources \citep{2015A&A...576A..38B}, constituting the low-luminosity bulk of the radio emitting AGN population, are thought to give a contribution of up to $\sim$20\% to the unresolved $\gamma$-ray background \citep{2019ApJ...879...68S}. In addition, recent \emph{Fermi}/LAT studies on the population of young radio sources \citep{2020A&A...635A.185P,2021MNRAS.507.4564P} found that some individual objects can be bright $\gamma$-ray sources, although the population as a whole seems to lie below the \emph{Fermi}/LAT detection threshold. The compact nature of these sources will be better investigated with future radio surveys, as well as with future higher resolution $\gamma$-ray observatories, possibly revealing a larger population of radio galaxies emitting at GeV energies.


\subsection{GeV emission in the \emph{INTEGRAL} hard X-ray AGN sample} 

IGR\,J18249--3243 is included in the \emph{INTEGRAL} complete sample of AGN \citep{2012MNRAS.426.1750M}, which is now fully covered by \emph{Fermi}/LAT observations, providing an  idea on the role of gamma ray emission in local hard X-ray bright AGN. In particular, if compared to the latest \emph{Fermi} 10 year catalogue, this sample can give an estimate of how many local AGN emit strongly above 100-200 keV, reaching the GeV band. The sample  was extracted from the third \emph{INTEGRAL}/IBIS survey to study the distribution of the absorption in the local Universe \citep{2012MNRAS.426.1750M}. It consists of  87 objects \footnote{one source, IGR J03184-0014, is not considered here as it was never seen again in subsequent INTEGRAL surveys} divided in 46 type 1 (Seyfert 1-1.5, of which 5 Narrow Line Seyfert 1) and 33 type 2 (Seyfert 1.8-2) AGN; only 8 blazars (BLLac-QSO) were reported in the catalogue. All of  these 8 blazars have emission in the MeV/GeV band, confirming that despite being rare they belong to the main contributing class at high energies. 
Another object with GeV detection in this complete INTEGRAL  sample of AGN (IGR\,J13109--5552) is classified as a blazar candidate of uncertain type; in order to 
define its true nature we obtained ATCA radio observations and also analysed measurements available in the literature: as a consequence  we have been able to classify the source as a true blazar of the Flat Spectrum Radio Quasar type (see Appendix \ref{1310}).
This emphasised once more that hard X-ray emitting blazars emit up to the GeV domain.

Among the Seyfert population, radio galaxies with the typical double-lobed morphology are also GeV emitters: before this work, 4 objects (NGC\,1275, B3\,0309+411B, 3C\,111 and Cen\,A) belonging  to the sample  were known to be detected by \emph{Fermi}/LAT.  With the identification of IGR\,J18249--3243 as a new radio galaxy, we bring the above number to 5. 

Finally, 3 Seyferts of type 2 (Circinus, NGC 1068 and NGC 4945) have also been detected at GeV energies: interestingly all of them are Compton thick objects (i.e. heavily absorbed), but their $\gamma$-ray emission is probably associated to strong starburst activity in their host galaxy. 

Overall,  the gamma ray detection rate within the INTEGRAL complete sample of AGN  is  20$\pm$5$\%$, emphasising the fact that in the majority of local active galaxies (80$\%$) the primary continuum drops off exponentially above a few hundred keV. Besides blazars, the only other exceptions are radio jetted AGN and possibly objects with strong starburst activity. Through this work we also found that, out of 11 radio galaxies present in the INTEGRAL sample, 5 (or 45$\%$) have a detection in the \emph{Fermi} 10th year catalogue; the next step will be to  understand what makes these 5 objects peculiar with respect to the other 6 which do not display GeV emission.


\section{Conclusions}

We can summarise our findings as follows:
\begin{itemize}
    \item Thanks to the recent release of the first epoch of the VLASS survey, we identified a new \emph{INTEGRAL} FR II radio galaxy with detection at GeV energies by \emph{Fermi}/LAT.
    \item Based on broad band SED modeling, we found that jet contribution is not sufficient to reproduce the GeV emission. Instead, we showed how it could be accounted for when considering the emission of lobes, and in particular the inverse Compton component from radio-emitting electron off ambient photon fields. This result stresses the contribution of broad-band SED modeling in characterizing the different emission regions and mechanisms, overcoming the lack of resolution at high energies.
    \item Thanks to data from the new generation radio surveys VLASS, RACS, and LoTSS, we present a list of four newly discovered radio galaxies detected by \emph{Fermi}/LAT. Additional four sources were compiled from the literature, for a total of eight objects. Despite their smaller projected linear size, their overall $\gamma$-ray properties are similar to the ones of previously known similar objects, suggesting the same preferred polar orientation.
    \item We foresee that further GeV emitting radio galaxies will be found thanks to new radio surveys, providing a deeper and sharper view on the counterparts of the GeV sky. This will unveil an emerging population of radio galaxies, in addition to the larger population of blazars commonly found, making possible the study of $\gamma$-ray emission from their extended regions as done in this paper for the lobes of IGR\,J18249--3243.
\end{itemize}


\section*{Acknowledgements}
 The authors acknowledge financial support from ASI under contract n. 2019-35-HH.0, in particular for G.B. research contract. The Australia Telescope Compact Array is part of the Australia Telescope National Facility (grid.421683.a) which is funded by the Australian Government for operation as a National Facility managed by CSIRO. The National Radio Astronomy Observatory is a facility of the National Science Foundation operated under cooperative agreement by Associated Universities, Inc. CIRADA is funded by a grant from the Canada Foundation for Innovation 2017 Innovation Fund (Project 35999), as well as by the Provinces of Ontario, British Columbia, Alberta, Manitoba and Quebec. The Pan-STARRS1  Surveys  (PS1)  and  the  PS1  public  science  archive have  been  made  possible  through  contributions  by  the  Institute for Astronomy, the University of Hawaii, the Pan-STARRS Project Office, the Max Planck Society and its participating institutes, the Max Planck Institute for Astronomy, Heidelberg and the Max Planck Institute for Extraterrestrial Physics, Garching, the Johns Hopkins University,  Durham  University,  the  University  of  Edinburgh,  theQueen’s University Belfast, the Harvard-Smithsonian Center for Astrophysics, the Las Cumbres Observatory Global Telescope Network Incorporated, the National Central University of Taiwan, the Space Telescope  Science  Institute,  the  National  Aeronautics  and  Space Administration under Grant No. NNX08AR22G issued through the Planetary  Science  Division  of  the  NASA  Science  Mission  Directorate,  the  National  Science  Foundation  Grant  No.  AST-1238877,the University of Maryland, Eotvos Lorand University (ELTE), the Los Alamos National Laboratory, and the Gordon and Betty Moore Foundation. This research has made use of the VizieR catalogue access tool, CDS, Strasbourg, France (DOI: 10.26093/cds/vizier). LOFAR data products were provided by the LOFAR Surveys Key Science project (LSKSP; https://lofar-surveys.org/) and were derived from observations with the International LOFAR Telescope (ILT). LOFAR (van Haarlem et al. 2013) is the Low Frequency Array designed and constructed by ASTRON. It has observing, data processing, and data storage facilities in several countries, which are owned by various parties (each with their own funding sources), and which are collectively operated by the ILT foundation under a joint scientific policy. The efforts of the LSKSP have benefited from funding from the European Research Council, NOVA, NWO, CNRS-INSU, the SURF Co-operative, the UK Science and Technology Funding Council and the Jülich Supercomputing Centre.
 
 
 \section*{Data availability}
Most of data presented in this work are public and available from relevant archives. Those not yet public will be made available from the corresponding author upon request.
 
\bibliographystyle{mnras}
\bibliography{grg-biblio} 

\begin{thebibliography}{}
\makeatletter
\relax
\def\mn@urlcharsother{\let\do\@makeother \do\$\do\&\do\#\do\^\do\_\do\%\do\~}
\def\mn@doi{\begingroup\mn@urlcharsother \@ifnextchar [ {\mn@doi@}
  {\mn@doi@[]}}
\def\mn@doi@[#1]#2{\def\@tempa{#1}\ifx\@tempa\@empty \href
  {http://dx.doi.org/#2} {doi:#2}\else \href {http://dx.doi.org/#2} {#1}\fi
  \endgroup}
\def\mn@eprint#1#2{\mn@eprint@#1:#2::\@nil}
\def\mn@eprint@arXiv#1{\href {http://arxiv.org/abs/#1} {{\tt arXiv:#1}}}
\def\mn@eprint@dblp#1{\href {http://dblp.uni-trier.de/rec/bibtex/#1.xml}
  {dblp:#1}}
\def\mn@eprint@#1:#2:#3:#4\@nil{\def\@tempa {#1}\def\@tempb {#2}\def\@tempc
  {#3}\ifx \@tempc \@empty \let \@tempc \@tempb \let \@tempb \@tempa \fi \ifx
  \@tempb \@empty \def\@tempb {arXiv}\fi \@ifundefined
  {mn@eprint@\@tempb}{\@tempb:\@tempc}{\expandafter \expandafter \csname
  mn@eprint@\@tempb\endcsname \expandafter{\@tempc}}}

\bibitem[\protect\citeauthoryear{{Abdo} et~al.,}{{Abdo}
  et~al.}{2010a}]{2010Sci...328..725A}
{Abdo} A.~A.,  et~al., 2010a, \mn@doi [Science] {10.1126/science.1184656},
  \href {https://ui.adsabs.harvard.edu/abs/2010Sci...328..725A} {328, 725}

\bibitem[\protect\citeauthoryear{{Abdo} et~al.,}{{Abdo}
  et~al.}{2010b}]{2010ApJ...720..912A}
{Abdo} A.~A.,  et~al., 2010b, \mn@doi [\apj] {10.1088/0004-637X/720/1/912},
  \href {https://ui.adsabs.harvard.edu/abs/2010ApJ...720..912A} {720, 912}

\bibitem[\protect\citeauthoryear{{Abdollahi} et~al.,}{{Abdollahi}
  et~al.}{2020}]{2020ApJS..247...33A}
{Abdollahi} S.,  et~al., 2020, \mn@doi [\apjs] {10.3847/1538-4365/ab6bcb},
  \href {https://ui.adsabs.harvard.edu/abs/2020ApJS..247...33A} {247, 33}

\bibitem[\protect\citeauthoryear{{Acero} et~al.,}{{Acero}
  et~al.}{2015}]{2015ApJS..218...23A}
{Acero} F.,  et~al., 2015, \mn@doi [\apjs] {10.1088/0067-0049/218/2/23}, \href
  {https://ui.adsabs.harvard.edu/abs/2015ApJS..218...23A} {218, 23}

\bibitem[\protect\citeauthoryear{{Ackermann} et~al.,}{{Ackermann}
  et~al.}{2011}]{2011ApJ...741...30A}
{Ackermann} M.,  et~al., 2011, \mn@doi [\apj] {10.1088/0004-637X/741/1/30},
  \href {https://ui.adsabs.harvard.edu/abs/2011ApJ...741...30A} {741, 30}

\bibitem[\protect\citeauthoryear{{Ackermann} et~al.,}{{Ackermann}
  et~al.}{2016}]{2016ApJ...826....1A}
{Ackermann} M.,  et~al., 2016, \mn@doi [\apj] {10.3847/0004-637X/826/1/1},
  \href {https://ui.adsabs.harvard.edu/abs/2016ApJ...826....1A} {826, 1}

\bibitem[\protect\citeauthoryear{{Ajello} et~al.,}{{Ajello}
  et~al.}{2020}]{2020ApJ...892..105A}
{Ajello} M.,  et~al., 2020, \mn@doi [\apj] {10.3847/1538-4357/ab791e}, \href
  {https://ui.adsabs.harvard.edu/abs/2020ApJ...892..105A} {892, 105}

\bibitem[\protect\citeauthoryear{{Angioni}}{{Angioni}}{2020}]{2020APh...11602393A}
{Angioni} R.,  2020, \mn@doi [Astroparticle Physics]
  {10.1016/j.astropartphys.2019.102393}, \href
  {https://ui.adsabs.harvard.edu/abs/2020APh...11602393A} {116, 102393}

\bibitem[\protect\citeauthoryear{{Baldi}, {Capetti}  \& {Giovannini}}{{Baldi}
  et~al.}{2015}]{2015A&A...576A..38B}
{Baldi} R.~D.,  {Capetti} A.,   {Giovannini} G.,  2015, \mn@doi [\aap]
  {10.1051/0004-6361/201425426}, \href
  {https://ui.adsabs.harvard.edu/abs/2015A&A...576A..38B} {576, A38}

\bibitem[\protect\citeauthoryear{{Bassani}, {Venturi}, {Molina}, {Malizia},
  {Dallacasa}, {Panessa}, {Bazzano}  \& {Ubertini}}{{Bassani}
  et~al.}{2016}]{2016MNRAS.461.3165B}
{Bassani} L.,  {Venturi} T.,  {Molina} M.,  {Malizia} A.,  {Dallacasa} D.,
  {Panessa} F.,  {Bazzano} A.,   {Ubertini} P.,  2016, \mn@doi [\mnras]
  {10.1093/mnras/stw1468}, \href
  {https://ui.adsabs.harvard.edu/abs/2016MNRAS.461.3165B} {461, 3165}

\bibitem[\protect\citeauthoryear{{Bird} et~al.,}{{Bird}
  et~al.}{2007}]{2007ApJS..170..175B}
{Bird} A.~J.,  et~al., 2007, \mn@doi [\apjs] {10.1086/513148}, \href
  {https://ui.adsabs.harvard.edu/abs/2007ApJS..170..175B} {170, 175}

\bibitem[\protect\citeauthoryear{{Bird} et~al.,}{{Bird}
  et~al.}{2016}]{2016ApJS..223...15B}
{Bird} A.~J.,  et~al., 2016, \mn@doi [\apjs] {10.3847/0067-0049/223/1/15},
  \href {https://ui.adsabs.harvard.edu/abs/2016ApJS..223...15B} {223, 15}

\bibitem[\protect\citeauthoryear{{Buzzoni}, {Arnaboldi}  \&
  {Corradi}}{{Buzzoni} et~al.}{2006}]{2006MNRAS.368..877B}
{Buzzoni} A.,  {Arnaboldi} M.,   {Corradi} R. L.~M.,  2006, \mn@doi [\mnras]
  {10.1111/j.1365-2966.2006.10163.x}, \href
  {https://ui.adsabs.harvard.edu/abs/2006MNRAS.368..877B} {368, 877}

\bibitem[\protect\citeauthoryear{{Chiaro}, {La Mura}, {Dom{\'\i}nguez}  \&
  {Bisogni}}{{Chiaro} et~al.}{2020}]{2020JHEAp..27...77C}
{Chiaro} G.,  {La Mura} G.,  {Dom{\'\i}nguez} A.,   {Bisogni} S.,  2020,
  \mn@doi [Journal of High Energy Astrophysics] {10.1016/j.jheap.2020.07.002},
  \href {https://ui.adsabs.harvard.edu/abs/2020JHEAp..27...77C} {27, 77}

\bibitem[\protect\citeauthoryear{{Cotton} et~al.,}{{Cotton}
  et~al.}{2020}]{2020MNRAS.495.1271C}
{Cotton} W.~D.,  et~al., 2020, \mn@doi [\mnras] {10.1093/mnras/staa1240}, \href
  {https://ui.adsabs.harvard.edu/abs/2020MNRAS.495.1271C} {495, 1271}

\bibitem[\protect\citeauthoryear{{D'Ammando} et~al.,}{{D'Ammando}
  et~al.}{2015}]{2015MNRAS.450.3975D}
{D'Ammando} F.,  et~al., 2015, \mn@doi [\mnras] {10.1093/mnras/stv909}, \href
  {https://ui.adsabs.harvard.edu/abs/2015MNRAS.450.3975D} {450, 3975}

\bibitem[\protect\citeauthoryear{{Foschini} et~al.,}{{Foschini}
  et~al.}{2021}]{2021arXivF}
{Foschini} L.,  et~al., 2021, arXiv e-prints, \href
  {https://ui.adsabs.harvard.edu/abs/2021arXiv211001995F} {p. arXiv:2110.01995}

\bibitem[\protect\citeauthoryear{{Franceschini} \& {Rodighiero}}{{Franceschini}
  \& {Rodighiero}}{2017}]{2017A&A...603A..34F}
{Franceschini} A.,  {Rodighiero} G.,  2017, \mn@doi [\aap]
  {10.1051/0004-6361/201629684}, \href
  {https://ui.adsabs.harvard.edu/abs/2017A&A...603A..34F} {603, A34}

\bibitem[\protect\citeauthoryear{{Giroletti} \& {Polatidis}}{{Giroletti} \&
  {Polatidis}}{2009}]{2009AN....330..193G}
{Giroletti} M.,  {Polatidis} A.,  2009, \mn@doi [Astronomische Nachrichten]
  {10.1002/asna.200811154}, \href
  {https://ui.adsabs.harvard.edu/abs/2009AN....330..193G} {330, 193}

\bibitem[\protect\citeauthoryear{{Grandi} \& {Palumbo}}{{Grandi} \&
  {Palumbo}}{2007}]{2007ApJ...659..235G}
{Grandi} P.,  {Palumbo} G. G.~C.,  2007, \mn@doi [\apj] {10.1086/510769}, \href
  {https://ui.adsabs.harvard.edu/abs/2007ApJ...659..235G} {659, 235}

\bibitem[\protect\citeauthoryear{{Grandi}, {Torresi}  \&
  {Stanghellini}}{{Grandi} et~al.}{2012}]{2012ApJ...751L...3G}
{Grandi} P.,  {Torresi} E.,   {Stanghellini} C.,  2012, \mn@doi [\apjl]
  {10.1088/2041-8205/751/1/L3}, \href
  {https://ui.adsabs.harvard.edu/abs/2012ApJ...751L...3G} {751, L3}

\bibitem[\protect\citeauthoryear{{Griffith} \& {Wright}}{{Griffith} \&
  {Wright}}{1993}]{1993AJ....105.1666G}
{Griffith} M.~R.,  {Wright} A.~E.,  1993, \mn@doi [\aj] {10.1086/116545}, \href
  {https://ui.adsabs.harvard.edu/abs/1993AJ....105.1666G} {105, 1666}

\bibitem[\protect\citeauthoryear{{Hern{\'a}ndez-Garc{\'\i}a}
  et~al.,}{{Hern{\'a}ndez-Garc{\'\i}a} et~al.}{2017}]{2017AA...603A.131H}
{Hern{\'a}ndez-Garc{\'\i}a} L.,  et~al., 2017, \mn@doi [\aap]
  {10.1051/0004-6361/201730530}, \href
  {https://ui.adsabs.harvard.edu/abs/2017A&A...603A.131H} {603, A131}

\bibitem[\protect\citeauthoryear{{Hern{\'a}ndez-Garc{\'\i}a}
  et~al.,}{{Hern{\'a}ndez-Garc{\'\i}a} et~al.}{2018}]{2018MNRAS.478.4634H}
{Hern{\'a}ndez-Garc{\'\i}a} L.,  et~al., 2018, \mn@doi [\mnras]
  {10.1093/mnras/sty1345}, \href
  {https://ui.adsabs.harvard.edu/abs/2018MNRAS.478.4634H} {478, 4634}

\bibitem[\protect\citeauthoryear{{Intema}, {Jagannathan}, {Mooley}  \&
  {Frail}}{{Intema} et~al.}{2017}]{2017A&A...598A..78I}
{Intema} H.~T.,  {Jagannathan} P.,  {Mooley} K.~P.,   {Frail} D.~A.,  2017,
  \mn@doi [\aap] {10.1051/0004-6361/201628536}, \href
  {https://ui.adsabs.harvard.edu/abs/2017A&A...598A..78I} {598, A78}

\bibitem[\protect\citeauthoryear{{Kataoka} et~al.,}{{Kataoka}
  et~al.}{2011}]{2011ApJ...740...29K}
{Kataoka} J.,  et~al., 2011, \mn@doi [\apj] {10.1088/0004-637X/740/1/29}, \href
  {https://ui.adsabs.harvard.edu/abs/2011ApJ...740...29K} {740, 29}

\bibitem[\protect\citeauthoryear{{Landi} et~al.,}{{Landi}
  et~al.}{2007}]{2007ATel.1273....1L}
{Landi} R.,  et~al., 2007, The Astronomer's Telegram, \href
  {https://ui.adsabs.harvard.edu/abs/2007ATel.1273....1L} {1273, 1}

\bibitem[\protect\citeauthoryear{{Landi} et~al.,}{{Landi}
  et~al.}{2009}]{2009A&A...493..893L}
{Landi} R.,  et~al., 2009, \mn@doi [\aap] {10.1051/0004-6361:200810503}, \href
  {https://ui.adsabs.harvard.edu/abs/2009A&A...493..893L} {493, 893}

\bibitem[\protect\citeauthoryear{{Lloyd} \& {Jones}}{{Lloyd} \&
  {Jones}}{2002}]{2002MNRAS.331..717L}
{Lloyd} B.~D.,  {Jones} P.~A.,  2002, \mn@doi [\mnras]
  {10.1046/j.1365-8711.2002.05239.x}, \href
  {https://ui.adsabs.harvard.edu/abs/2002MNRAS.331..717L} {331, 717}

\bibitem[\protect\citeauthoryear{{Lott}, {Gasparrini}  \& {Ciprini}}{{Lott}
  et~al.}{2020}]{2020arXiv201008406L}
{Lott} B.,  {Gasparrini} D.,   {Ciprini} S.,  2020, arXiv e-prints, \href
  {https://ui.adsabs.harvard.edu/abs/2020arXiv201008406L} {p. arXiv:2010.08406}

\bibitem[\protect\citeauthoryear{{Lubi{\'n}ski} et~al.,}{{Lubi{\'n}ski}
  et~al.}{2016}]{2016MNRAS.458.2454L}
{Lubi{\'n}ski} P.,  et~al., 2016, \mn@doi [\mnras] {10.1093/mnras/stw454},
  \href {https://ui.adsabs.harvard.edu/abs/2016MNRAS.458.2454L} {458, 2454}

\bibitem[\protect\citeauthoryear{{Malizia} et~al.,}{{Malizia}
  et~al.}{2007}]{2007ApJ...668...81M}
{Malizia} A.,  et~al., 2007, \mn@doi [\apj] {10.1086/520874}, \href
  {https://ui.adsabs.harvard.edu/abs/2007ApJ...668...81M} {668, 81}

\bibitem[\protect\citeauthoryear{{Malizia}, {Bassani}, {Bazzano}, {Bird},
  {Masetti}, {Panessa}, {Stephen}  \& {Ubertini}}{{Malizia}
  et~al.}{2012}]{2012MNRAS.426.1750M}
{Malizia} A.,  {Bassani} L.,  {Bazzano} A.,  {Bird} A.~J.,  {Masetti} N.,
  {Panessa} F.,  {Stephen} J.~B.,   {Ubertini} P.,  2012, \mn@doi [\mnras]
  {10.1111/j.1365-2966.2012.21755.x}, \href
  {https://ui.adsabs.harvard.edu/abs/2012MNRAS.426.1750M} {426, 1750}

\bibitem[\protect\citeauthoryear{{Malizia}, {Molina}, {Bassani}, {Stephen},
  {Bazzano}, {Ubertini}  \& {Bird}}{{Malizia}
  et~al.}{2014}]{2014ApJ...782L..25M}
{Malizia} A.,  {Molina} M.,  {Bassani} L.,  {Stephen} J.~B.,  {Bazzano} A.,
  {Ubertini} P.,   {Bird} A.~J.,  2014, \mn@doi [\apjl]
  {10.1088/2041-8205/782/2/L25}, \href
  {https://ui.adsabs.harvard.edu/abs/2014ApJ...782L..25M} {782, L25}

\bibitem[\protect\citeauthoryear{{Masetti} et~al.,}{{Masetti}
  et~al.}{2008}]{2008A&A...482..113M}
{Masetti} N.,  et~al., 2008, \mn@doi [\aap] {10.1051/0004-6361:20079332}, \href
  {https://ui.adsabs.harvard.edu/abs/2008A&A...482..113M} {482, 113}

\bibitem[\protect\citeauthoryear{{Masetti} et~al.,}{{Masetti}
  et~al.}{2009}]{2009A&A...495..121M}
{Masetti} N.,  et~al., 2009, \mn@doi [\aap] {10.1051/0004-6361:200811322},
  \href {https://ui.adsabs.harvard.edu/abs/2009A&A...495..121M} {495, 121}

\bibitem[\protect\citeauthoryear{{McKinley} et~al.,}{{McKinley}
  et~al.}{2015}]{2015MNRAS.446.3478M}
{McKinley} B.,  et~al., 2015, \mn@doi [\mnras] {10.1093/mnras/stu2310}, \href
  {https://ui.adsabs.harvard.edu/abs/2015MNRAS.446.3478M} {446, 3478}

\bibitem[\protect\citeauthoryear{{Migliori}, {Siemiginowska}, {Sobolewska},
  {Loh}, {Corbel}, {Ostorero}  \& {Stawarz}}{{Migliori}
  et~al.}{2016}]{2016ApJ...821L..31M}
{Migliori} G.,  {Siemiginowska} A.,  {Sobolewska} M.,  {Loh} A.,  {Corbel} S.,
  {Ostorero} L.,   {Stawarz} {\L}.,  2016, \mn@doi [\apjl]
  {10.3847/2041-8205/821/2/L31}, \href
  {https://ui.adsabs.harvard.edu/abs/2016ApJ...821L..31M} {821, L31}

\bibitem[\protect\citeauthoryear{{Miller} \& {Brandt}}{{Miller} \&
  {Brandt}}{2009}]{2009ApJ...695..755M}
{Miller} B.~P.,  {Brandt} W.~N.,  2009, \mn@doi [\apj]
  {10.1088/0004-637X/695/1/755}, \href
  {https://ui.adsabs.harvard.edu/abs/2009ApJ...695..755M} {695, 755}

\bibitem[\protect\citeauthoryear{{Morganti}, {Oosterloo}, {Tadhunter}, {Aiudi},
  {Jones}  \& {Villar-Martin}}{{Morganti} et~al.}{1999}]{1999AAS..140..355M}
{Morganti} R.,  {Oosterloo} T.,  {Tadhunter} C.~N.,  {Aiudi} R.,  {Jones} P.,
  {Villar-Martin} M.,  1999, \mn@doi [\aaps] {10.1051/aas:1999427}, \href
  {https://ui.adsabs.harvard.edu/abs/1999A&AS..140..355M} {140, 355}

\bibitem[\protect\citeauthoryear{{Onken} et~al.,}{{Onken}
  et~al.}{2019}]{2019PASA...36...33O}
{Onken} C.~A.,  et~al., 2019, \mn@doi [\pasa] {10.1017/pasa.2019.27}, \href
  {https://ui.adsabs.harvard.edu/abs/2019PASA...36...33O} {36, e033}

\bibitem[\protect\citeauthoryear{{Ostorero}, {Morganti}, {Diaferio},
  {Siemiginowska}, {Stawarz}, {Moderski}  \& {Labiano}}{{Ostorero}
  et~al.}{2017}]{2017ApJ...849...34O}
{Ostorero} L.,  {Morganti} R.,  {Diaferio} A.,  {Siemiginowska} A.,  {Stawarz}
  {\L}.,  {Moderski} R.,   {Labiano} A.,  2017, \mn@doi [\apj]
  {10.3847/1538-4357/aa8ef6}, \href
  {https://ui.adsabs.harvard.edu/abs/2017ApJ...849...34O} {849, 34}

\bibitem[\protect\citeauthoryear{{Paliya} et~al.,}{{Paliya}
  et~al.}{2019}]{2019ApJ...881..154P}
{Paliya} V.~S.,  et~al., 2019, \mn@doi [\apj] {10.3847/1538-4357/ab2f8b}, \href
  {https://ui.adsabs.harvard.edu/abs/2019ApJ...881..154P} {881, 154}

\bibitem[\protect\citeauthoryear{{Pe{\~n}a-Herazo} et~al.,}{{Pe{\~n}a-Herazo}
  et~al.}{2020}]{2020A&A...643A.103P}
{Pe{\~n}a-Herazo} H.~A.,  et~al., 2020, \mn@doi [\aap]
  {10.1051/0004-6361/202037978}, \href
  {https://ui.adsabs.harvard.edu/abs/2020A&A...643A.103P} {643, A103}

\bibitem[\protect\citeauthoryear{{Persic} \& {Rephaeli}}{{Persic} \&
  {Rephaeli}}{2007}]{2007A&A...463..481P}
{Persic} M.,  {Rephaeli} Y.,  2007, \mn@doi [\aap]
  {10.1051/0004-6361:20054146}, \href
  {https://ui.adsabs.harvard.edu/abs/2007A&A...463..481P} {463, 481}

\bibitem[\protect\citeauthoryear{{Persic} \& {Rephaeli}}{{Persic} \&
  {Rephaeli}}{2019a}]{2019MNRAS.485.2001P}
{Persic} M.,  {Rephaeli} Y.,  2019a, \mn@doi [\mnras] {10.1093/mnras/stz511},
  \href {https://ui.adsabs.harvard.edu/abs/2019MNRAS.485.2001P} {485, 2001}

\bibitem[\protect\citeauthoryear{{Persic} \& {Rephaeli}}{{Persic} \&
  {Rephaeli}}{2019b}]{2019MNRAS.490.1489P}
{Persic} M.,  {Rephaeli} Y.,  2019b, \mn@doi [\mnras] {10.1093/mnras/stz2527},
  \href {https://ui.adsabs.harvard.edu/abs/2019MNRAS.490.1489P} {490, 1489}

\bibitem[\protect\citeauthoryear{{Persic} \& {Rephaeli}}{{Persic} \&
  {Rephaeli}}{2020}]{2020MNRAS.491.5740P}
{Persic} M.,  {Rephaeli} Y.,  2020, \mn@doi [\mnras] {10.1093/mnras/stz3415},
  \href {https://ui.adsabs.harvard.edu/abs/2020MNRAS.491.5740P} {491, 5740}

\bibitem[\protect\citeauthoryear{{Prieto}, {Reunanen}, {Tristram}, {Neumayer},
  {Fernandez-Ontiveros}, {Orienti}  \& {Meisenheimer}}{{Prieto}
  et~al.}{2010}]{2010MNRAS.402..724P}
{Prieto} M.~A.,  {Reunanen} J.,  {Tristram} K.~R.~W.,  {Neumayer} N.,
  {Fernandez-Ontiveros} J.~A.,  {Orienti} M.,   {Meisenheimer} K.,  2010,
  \mn@doi [\mnras] {10.1111/j.1365-2966.2009.15897.x}, \href
  {https://ui.adsabs.harvard.edu/abs/2010MNRAS.402..724P} {402, 724}

\bibitem[\protect\citeauthoryear{{Principe} et~al.,}{{Principe}
  et~al.}{2020}]{2020A&A...635A.185P}
{Principe} G.,  et~al., 2020, \mn@doi [\aap] {10.1051/0004-6361/201937049},
  \href {https://ui.adsabs.harvard.edu/abs/2020A&A...635A.185P} {635, A185}

\bibitem[\protect\citeauthoryear{{Principe}, {Di Venere}, {Orienti},
  {Migliori}, {D'Ammando}, {Mazziotta}  \& {Giroletti}}{{Principe}
  et~al.}{2021}]{2021MNRAS.507.4564P}
{Principe} G.,  {Di Venere} L.,  {Orienti} M.,  {Migliori} G.,  {D'Ammando} F.,
   {Mazziotta} M.~N.,   {Giroletti} M.,  2021, \mn@doi [\mnras]
  {10.1093/mnras/stab2357}, \href
  {https://ui.adsabs.harvard.edu/abs/2021MNRAS.507.4564P} {507, 4564}

\bibitem[\protect\citeauthoryear{{Ricci} et~al.,}{{Ricci}
  et~al.}{2017}]{2017ApJS..233...17R}
{Ricci} C.,  et~al., 2017, \mn@doi [\apjs] {10.3847/1538-4365/aa96ad}, \href
  {https://ui.adsabs.harvard.edu/abs/2017ApJS..233...17R} {233, 17}

\bibitem[\protect\citeauthoryear{{Shimwell} et~al.,}{{Shimwell}
  et~al.}{2017}]{2017A&A...598A.104S}
{Shimwell} T.~W.,  et~al., 2017, \mn@doi [\aap] {10.1051/0004-6361/201629313},
  \href {https://ui.adsabs.harvard.edu/abs/2017A&A...598A.104S} {598, A104}

\bibitem[\protect\citeauthoryear{{Soldi} et~al.,}{{Soldi}
  et~al.}{2008}]{2008AA...486..411S}
{Soldi} S.,  et~al., 2008, \mn@doi [\aap] {10.1051/0004-6361:200809947}, \href
  {https://ui.adsabs.harvard.edu/abs/2008AA...486..411S} {486, 411}

\bibitem[\protect\citeauthoryear{{Stecker}, {Shrader}  \& {Malkan}}{{Stecker}
  et~al.}{2019}]{2019ApJ...879...68S}
{Stecker} F.~W.,  {Shrader} C.~R.,   {Malkan} M.~A.,  2019, \mn@doi [\apj]
  {10.3847/1538-4357/ab23ee}, \href
  {https://ui.adsabs.harvard.edu/abs/2019ApJ...879...68S} {879, 68}

\bibitem[\protect\citeauthoryear{{Torresi}, {Grandi}, {Capetti}, {Baldi}  \&
  {Giovannini}}{{Torresi} et~al.}{2018}]{2018MNRAS.476.5535T}
{Torresi} E.,  {Grandi} P.,  {Capetti} A.,  {Baldi} R.~D.,   {Giovannini} G.,
  2018, \mn@doi [\mnras] {10.1093/mnras/sty520}, \href
  {https://ui.adsabs.harvard.edu/abs/2018MNRAS.476.5535T} {476, 5535}

\bibitem[\protect\citeauthoryear{{Trinchieri}, {Wolter}  \&
  {Iovino}}{{Trinchieri} et~al.}{1998}]{1998LNP...506..551T}
{Trinchieri} G.,  {Wolter} A.,   {Iovino} A.,  1998, {A Ring of X-rays from the
  Cartwheel Galaxy}.
pp 551--554, \mn@doi{10.1007/BFb0104781}

\bibitem[\protect\citeauthoryear{{Wayth} et~al.,}{{Wayth}
  et~al.}{2015}]{2015PASA...32...25W}
{Wayth} R.~B.,  et~al., 2015, \mn@doi [\pasa] {10.1017/pasa.2015.26}, \href
  {https://ui.adsabs.harvard.edu/abs/2015PASA...32...25W} {32, e025}

\bibitem[\protect\citeauthoryear{{Wozniak}, {Zdziarski}, {Smith}, {Madejski}
  \& {Johnson}}{{Wozniak} et~al.}{1998}]{1998MNRAS.299..449W}
{Wozniak} P.~R.,  {Zdziarski} A.~A.,  {Smith} D.,  {Madejski} G.~M.,
  {Johnson} W.~N.,  1998, \mn@doi [\mnras] {10.1046/j.1365-8711.1998.01831.x},
  \href {https://ui.adsabs.harvard.edu/abs/1998MNRAS.299..449W} {299, 449}

\makeatother
\end{thebibliography}

\appendix

\section{Images of newly discovered GeV radio galaxies}

Here we report the images of the newly discovered radio galaxies from the latest radio surveys, discussed in Sec. \ref{sec:newRG}. For each source we report the VLASS radio image with RACS or LoTSS DR2 contours superimposed, and the overlay of radio contours on the optical image from Pan-STARRS or DSS2. The beam is reported in the lower-left corner. Contours are multiples of 3$\times$RMS, namely -1, 1, 2, 4, 8, 16, 32, 64, 128, 256. For sources 4FGL J0929.3--2414 and 4FGL J1455.4--3654 contours start from 12$\times$RMS, due to presence of image artifacts.


\begin{figure*}
\includegraphics[width=\columnwidth]{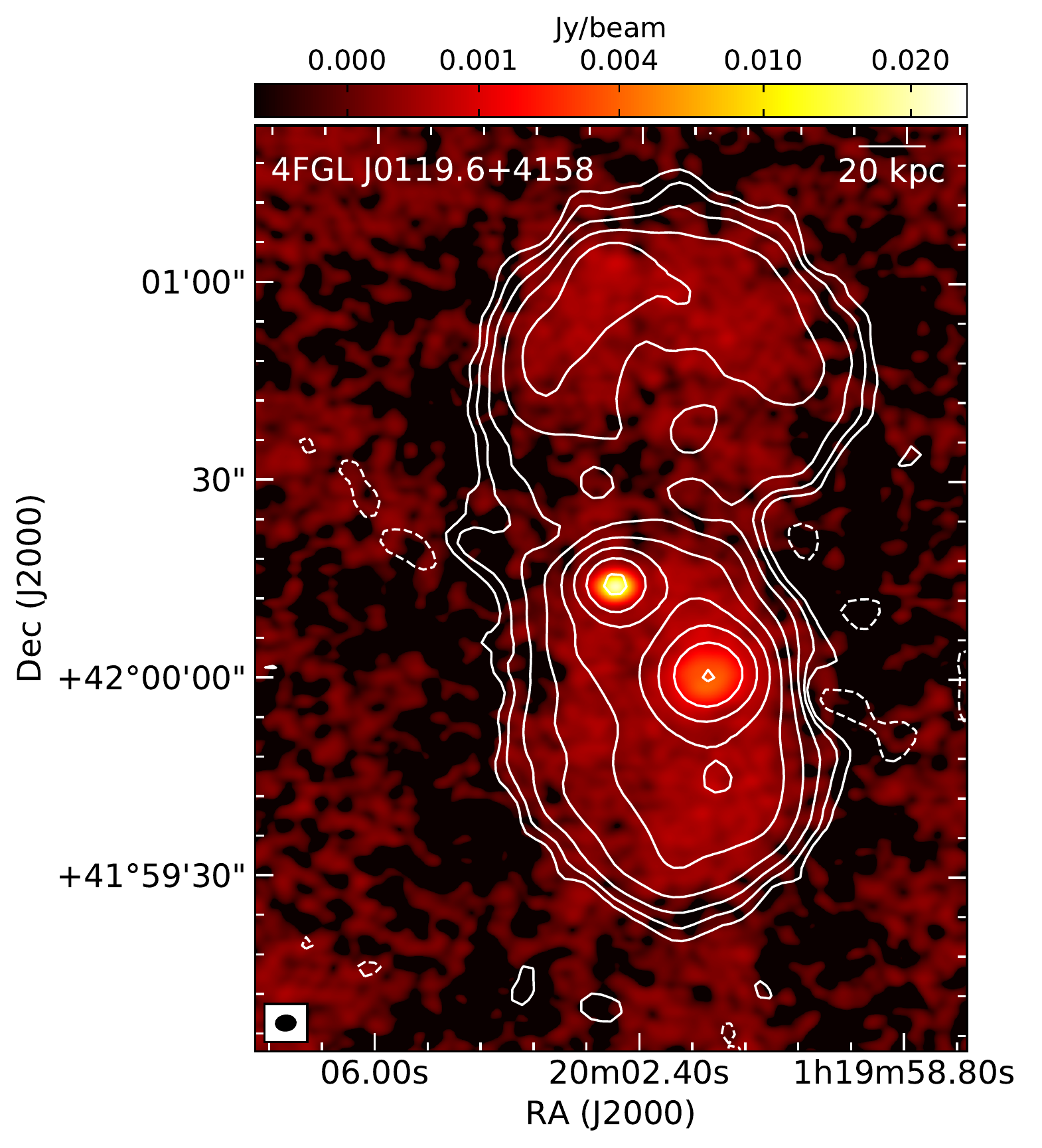}
\includegraphics[width=\columnwidth]{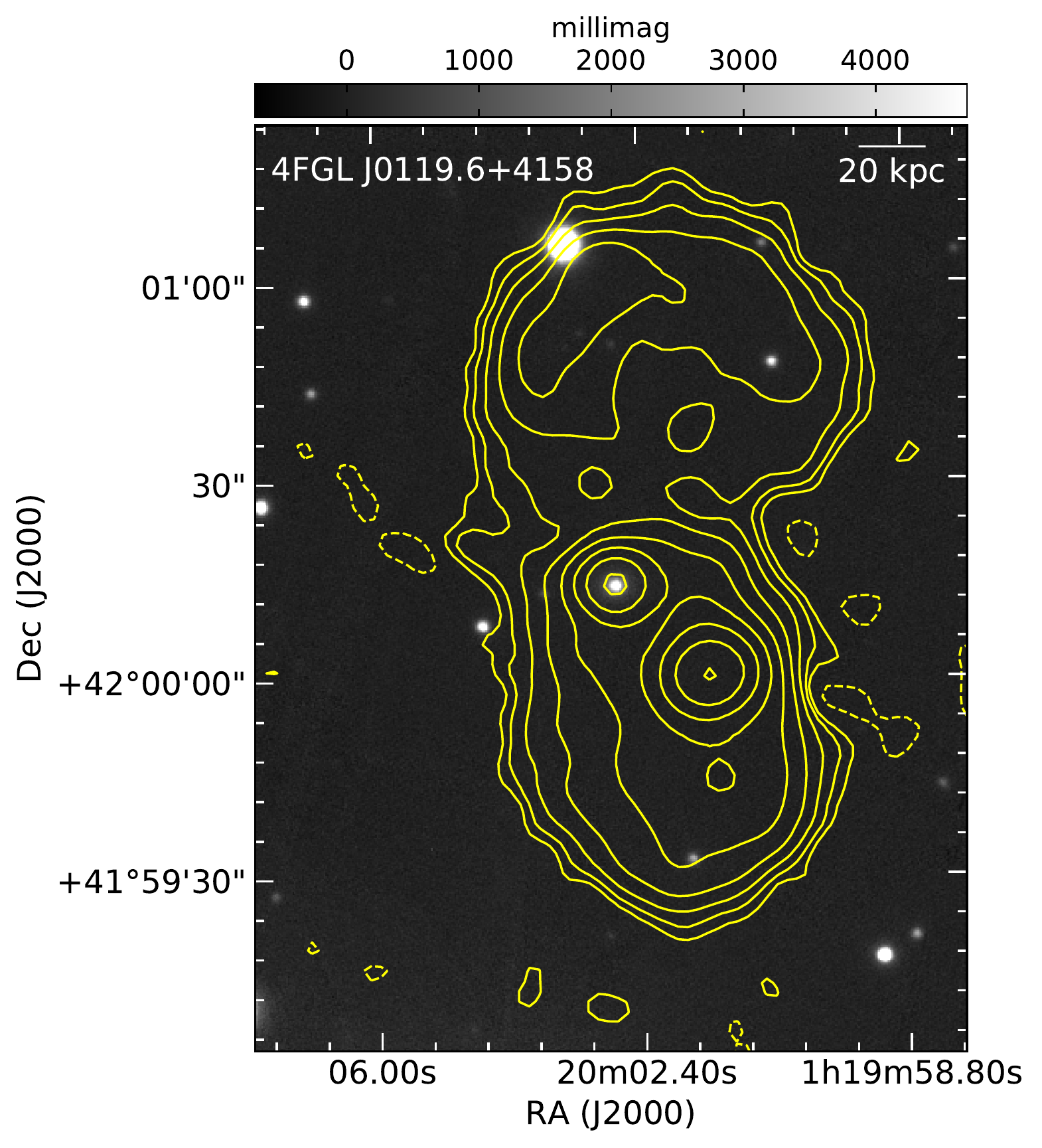}
\includegraphics[width=\columnwidth]{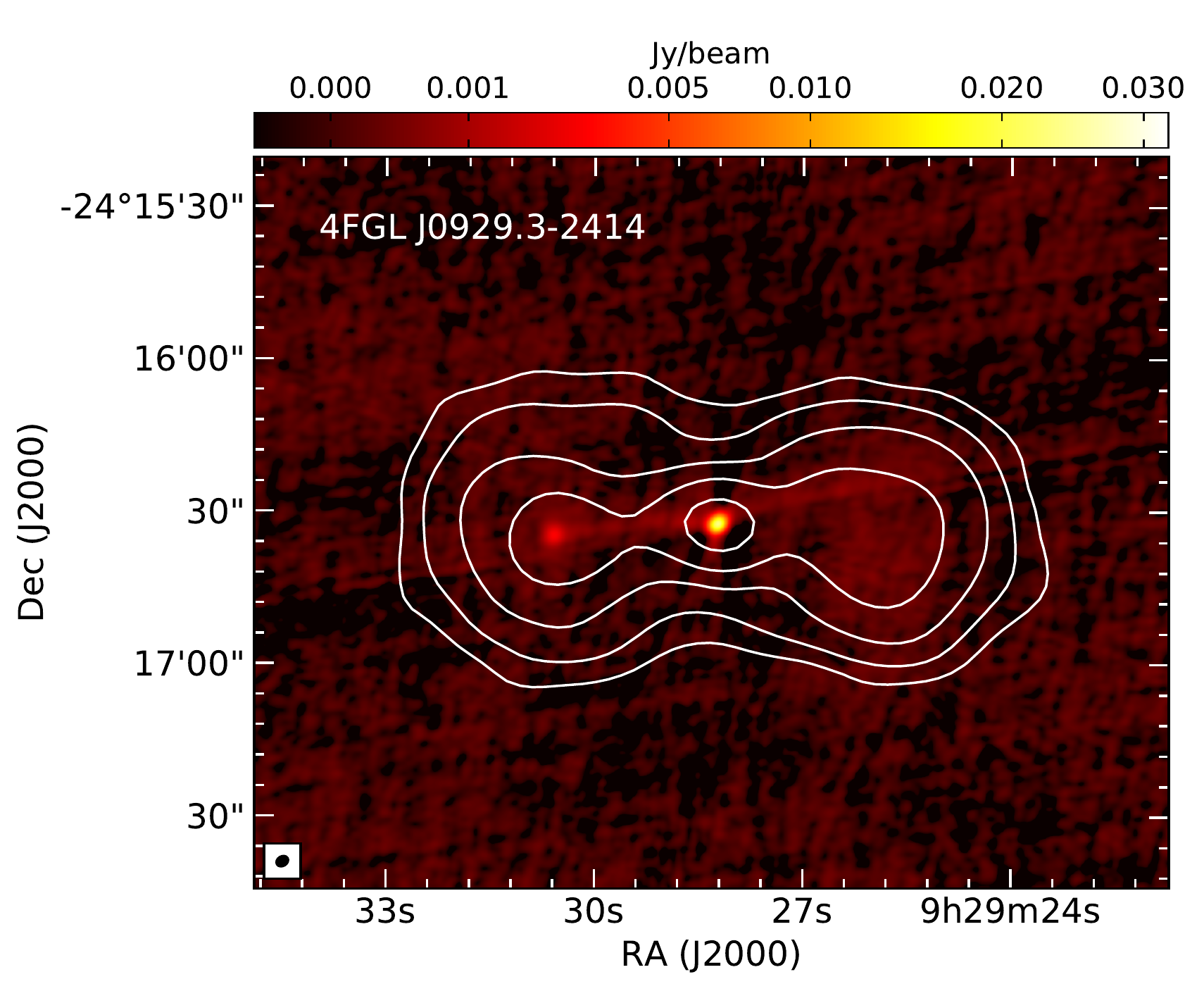}
\includegraphics[width=\columnwidth]{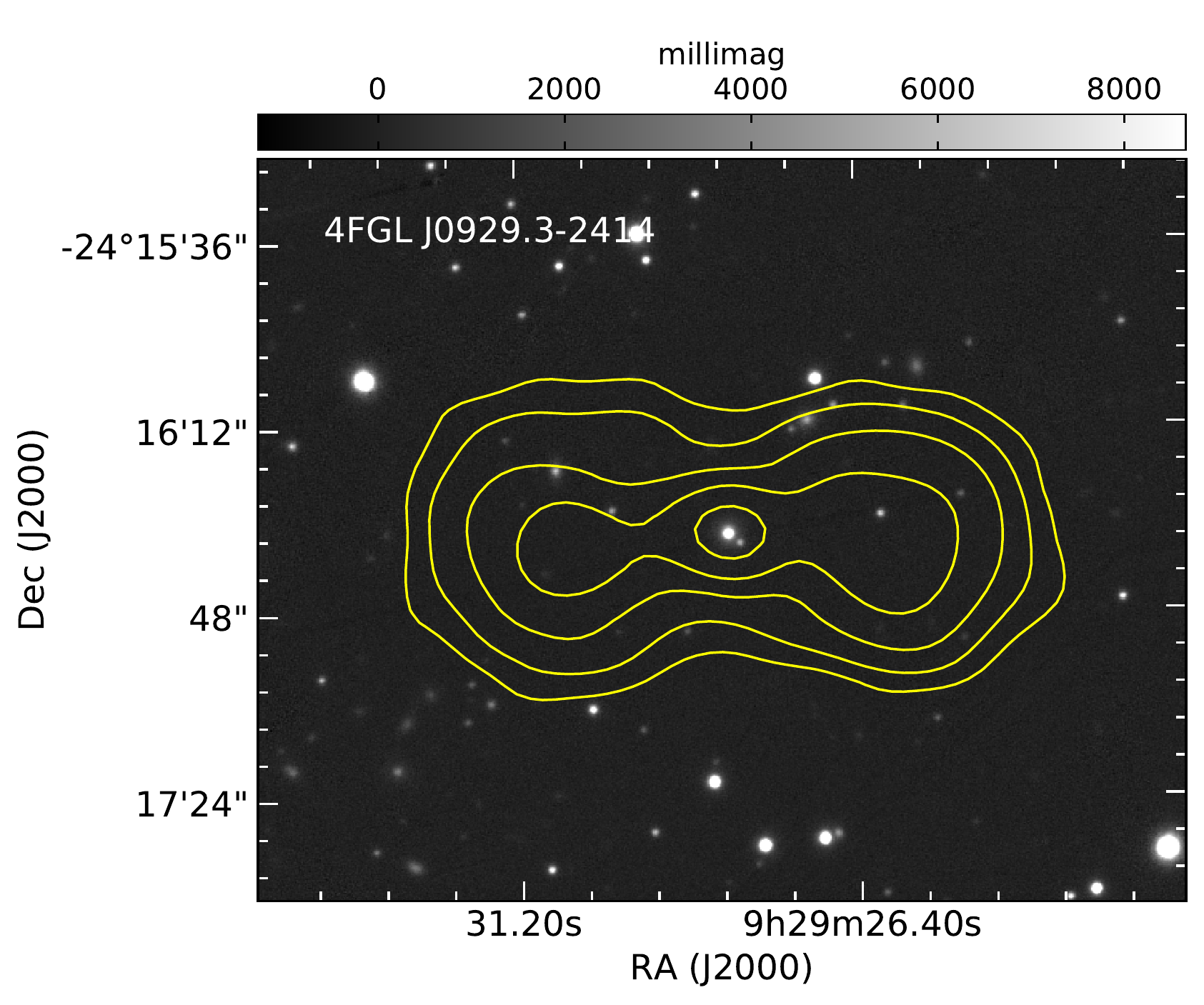}
 \caption{Images of newly discovered radio galaxies from recent surveys. Top panels: VLASS (3 GHz) image of source 4FGL\,J0119.6+4158 with LoTSS (150 MHz) contours superimposed (left) and Pan-STARRS image with LoTSS DR2 contours superimposed (right). Bottom panels: VLASS image of source 4FGL\,J0929.3-2414 with RACS (0.88 GHz) contours superimposed (left) and Pan-STARRS image with RACS contours superimposed (right).}
\label{MAGN1}
\end{figure*}


\begin{figure*}
\includegraphics[width=\columnwidth]{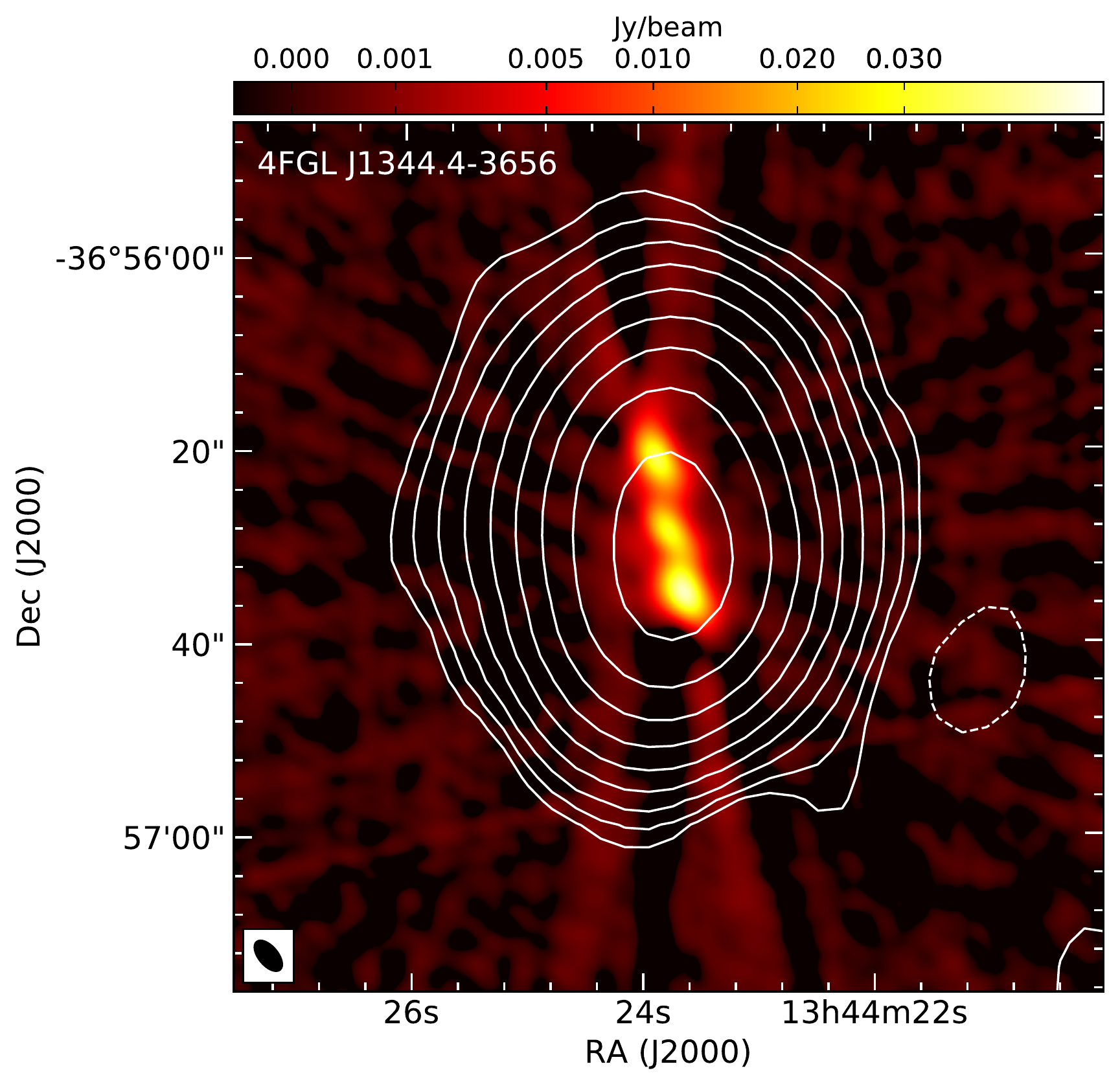}
\includegraphics[width=\columnwidth]{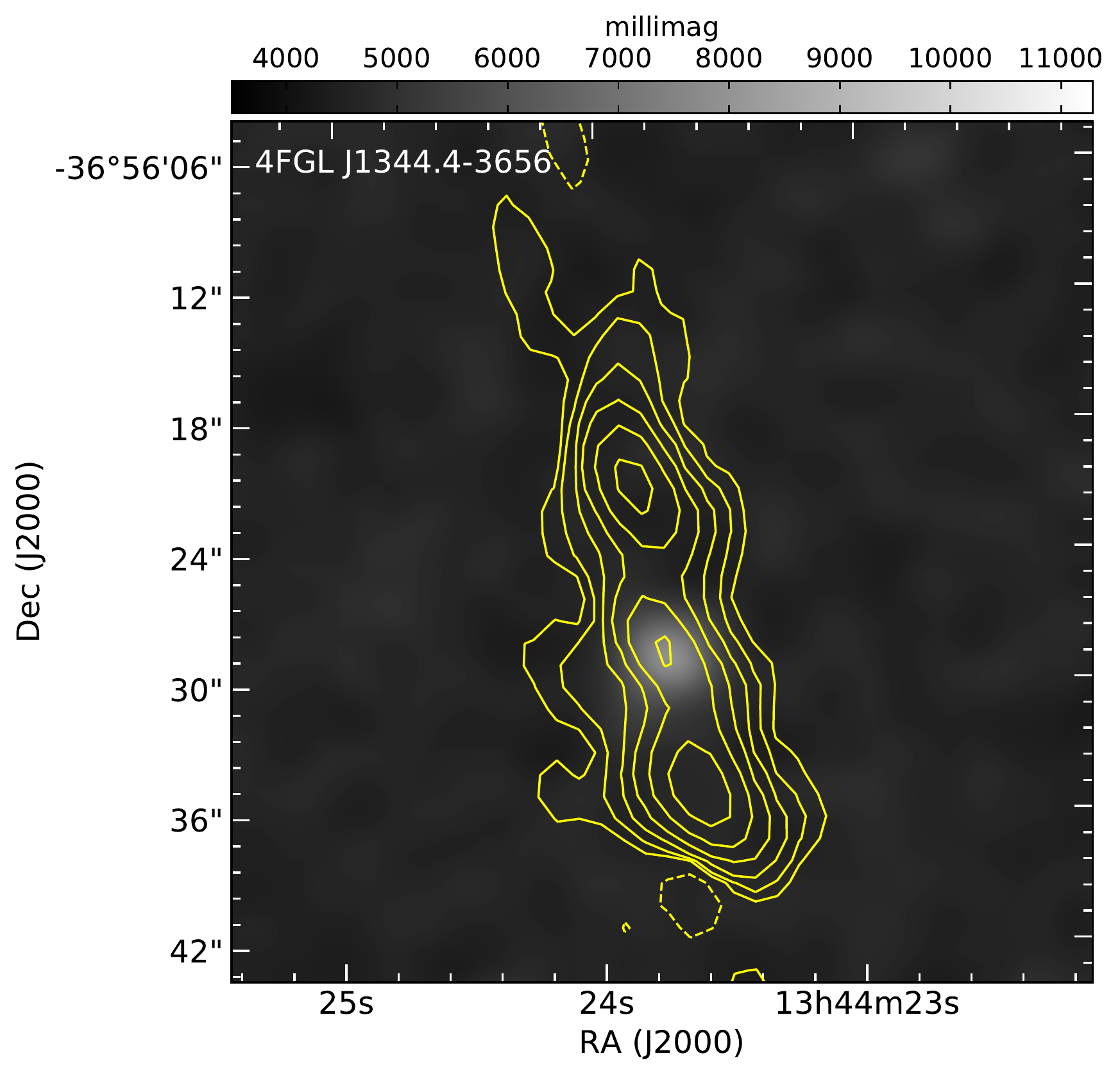}
\includegraphics[width=\columnwidth]{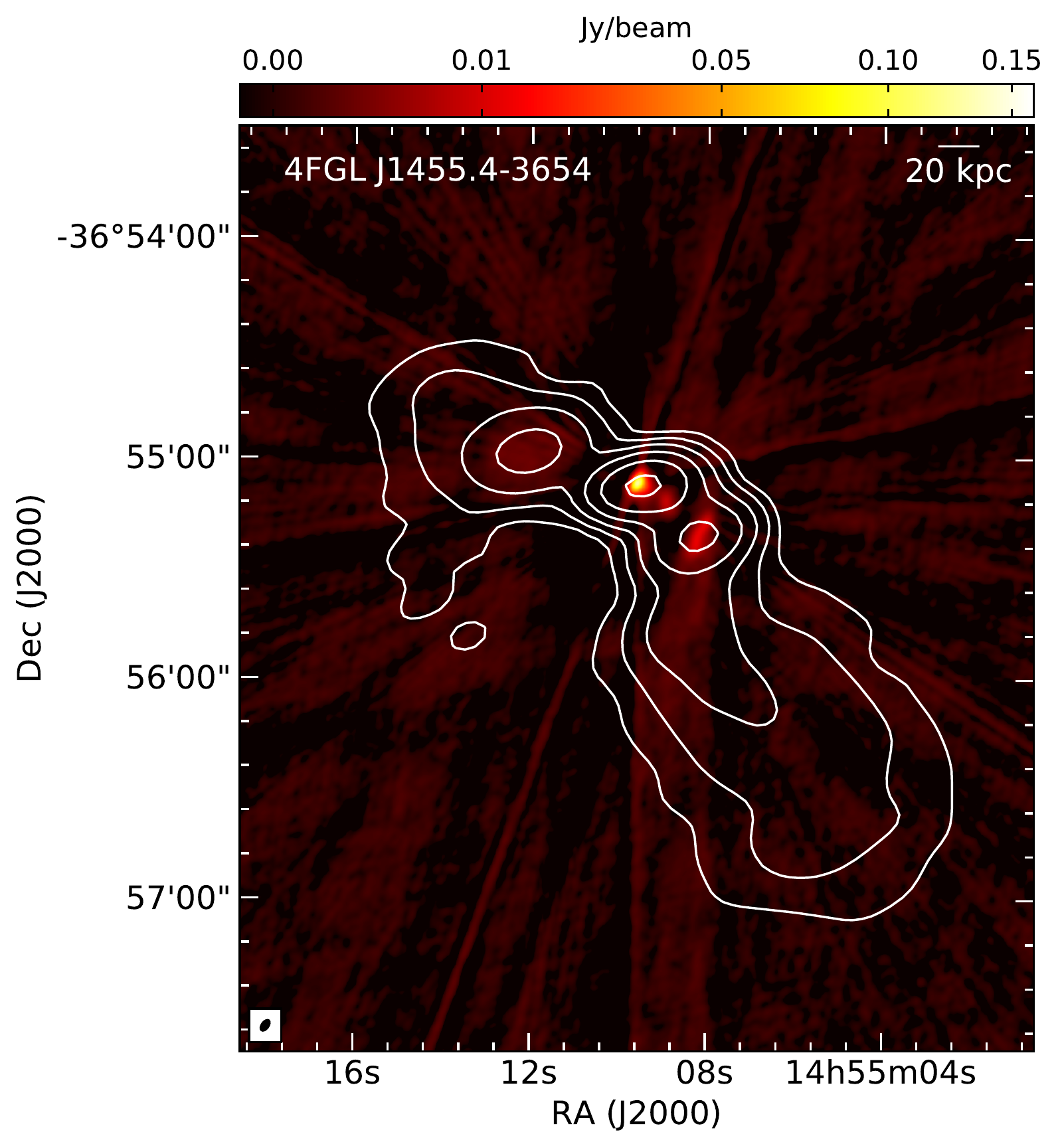}
\includegraphics[width=\columnwidth]{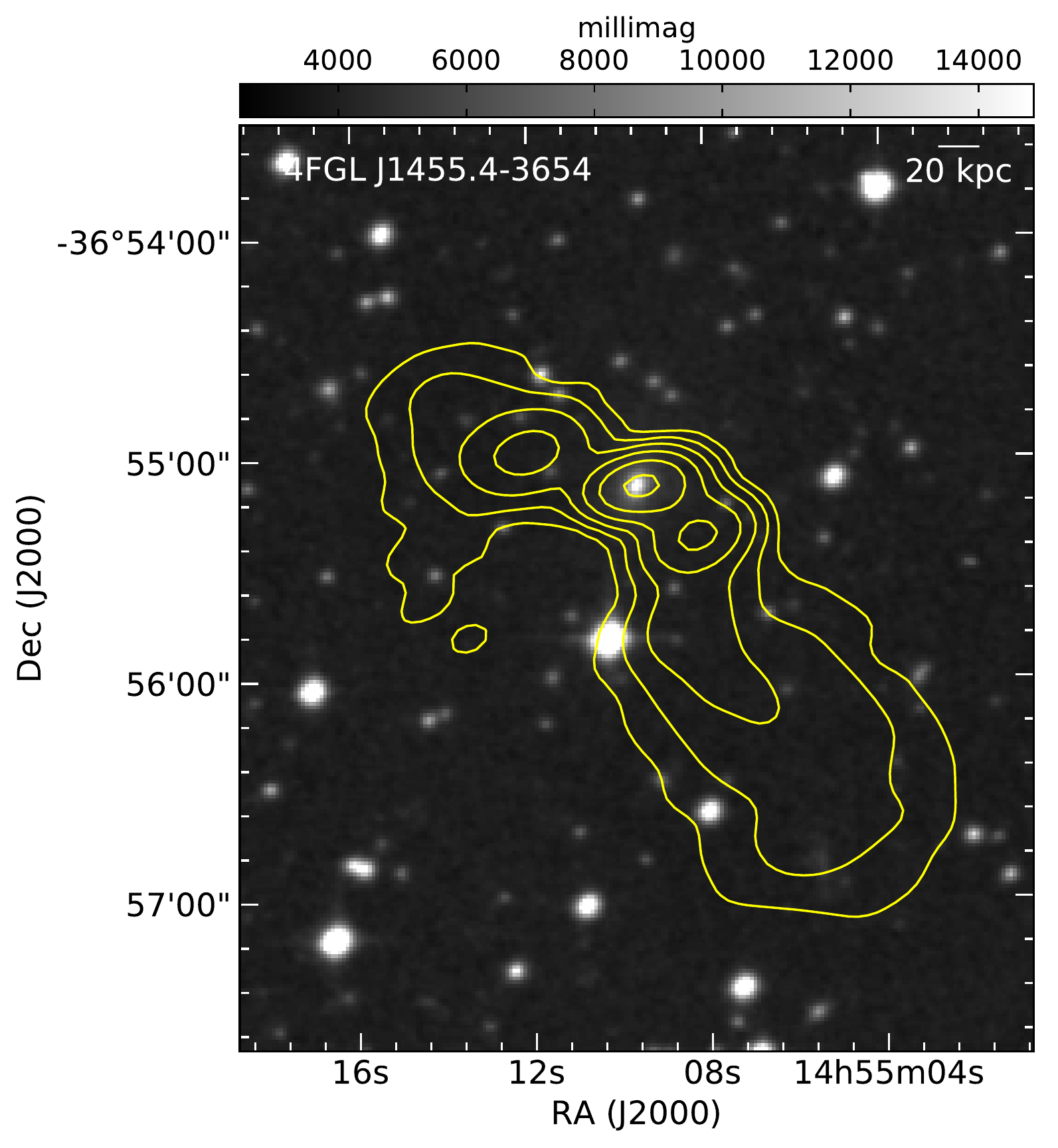}
 \caption{Same as for figure \ref{MAGN1}, but for sources 4FGL\,J1344.4-3656 and 4FGL\,J1455.4-3654. Top left panel is VLASS image with RACS contours superimposed, top right panel is the DSS2 image with VLASS contours superimposed. Bottom left panel is VLASS image with RACS contours, while bottom right panel DSS2 image with RACS contours.}
 
\label{MAGN2}
\end{figure*}


\section{The blazar IGR\,J13109--5552}
\label{1310}

This source was first reported by \emph{INTEGRAL}/IBIS as part of the third survey catalogue \citep{2007ApJS..170..175B}. Shortly after, \cite{2007ApJ...668...81M} provided the location of the X-ray counterpart by means of \emph{Swift}/XRT follow-up observations. The X-ray position at RA(J2000)=13:10:43.08, Dec(J2000)=--55:52:11.66 (positional uncertainty of 3.7 arcsec) allowed to match the source with a strong radio emitter (PMN\,J1310--5552). Furthermore, the restricted X-ray error circle enabled the identification of the optical counterpart, and through follow-up spectroscopic observations a loose \footnote{This is due to an anomalous optical spectrum, possibly caused by an intervening object. See \cite{2008A&A...482..113M} for more details} classification of the source: IGR\,J13109--5552 is a type 1 AGN  at a redshift of 0.104 \citep{2008A&A...482..113M}.  The source black hole mass is 2x10$^{8}$ solar masses \citep{2019ApJ...881..154P}.

The combined \emph{XMM}/pn plus \emph{INTEGRAL}/IBIS, and \emph{Swift}/XRT plus \emph{Swift}/BAT spectra, also presented by \cite{2014ApJ...782L..25M}, can be described by a simple power law ($\Gamma$=1.4-1.5) with no evidence of a high energy cut-off ($E_{cut}>360$ keV, \citealt{2014ApJ...782L..25M}) nor of reflection \citep{2017ApJS..233...17R}. The source is reported as persistent in the latest IBIS catalogue \citep{2016ApJS..223...15B}. In a recent analysis of beamed AGN detected by \emph{Swift}/BAT and \emph{Fermi}/LAT \citep{2019ApJ...881..154P}, it was classified as a blazar candidate of uncertain type due to its SED and radio detection. The source SED is indeed characterized by two broad peaks as observed in blazars, but that can also be found in some misaligned AGN. 

\begin{figure}
\includegraphics[width=\columnwidth]{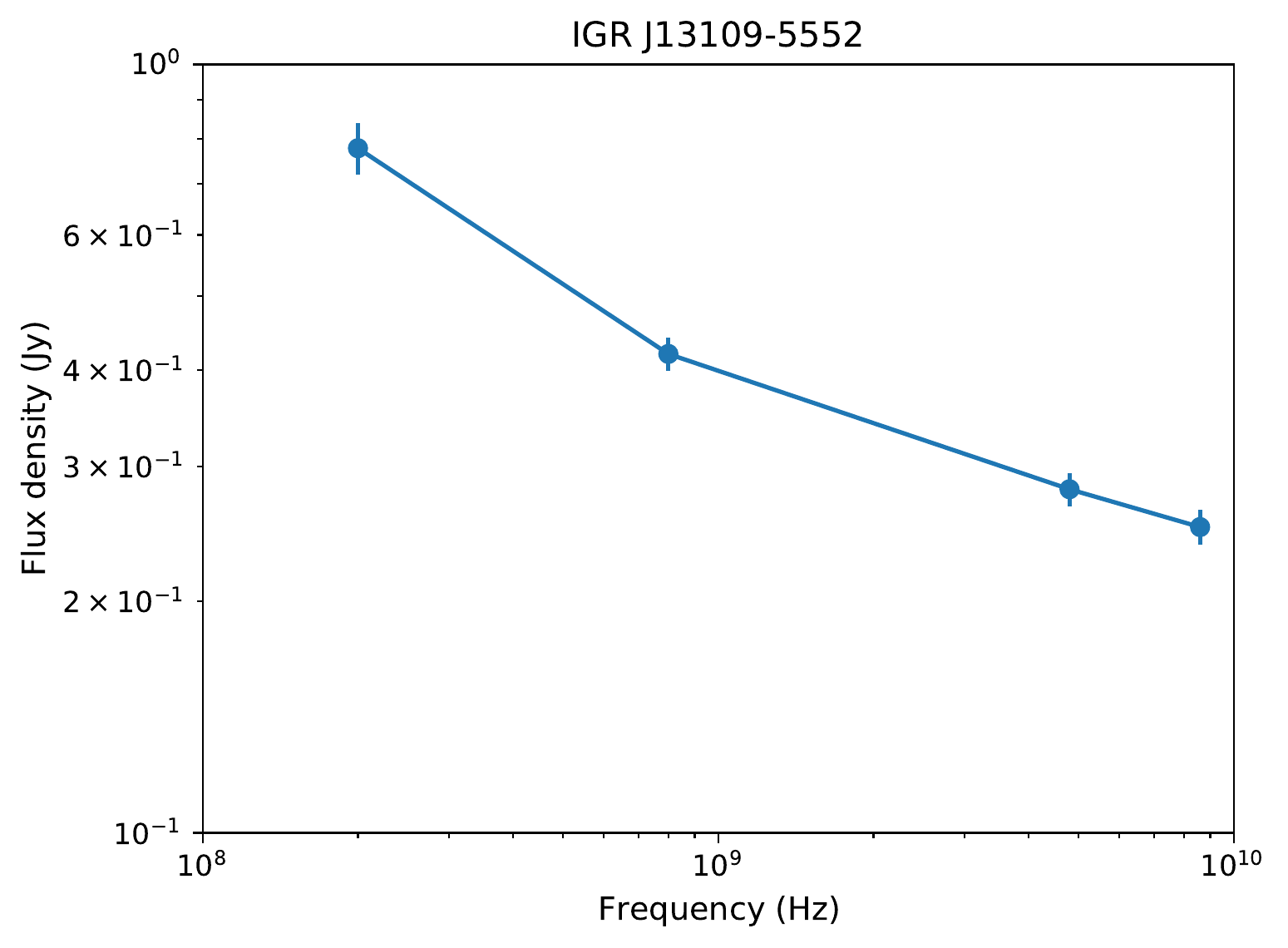}
 \caption{Radio SED for IGR\,J13109--5552 built with GLEAM (0.2 GHz), RACS (0.88 GHz) and ATCA data from our campaign (4.8, 8.6 GHz).}
\label{fig:1310}
\end{figure}

At radio wavelengths, IGR\,J13109--5552 is included in the most recent surveys of the Southern emisphere: GLEAM and RACS. In addition, we observed this source during the ATCA run described in Sec. \ref{sec:ATCA}, providing further two images at 4.8 and 8.6 GHz. The source results unresolved down to an angular resolution of 2.0\arcsec$\times$0.6\arcsec, provided by our highest resolution ATCA observations. At the redshift of the source, this translates into a projected linear size smaller than $\sim$1 kpc. This allows us to combine data from surveys and observations at different resolutions to build the radio SED of this object. The collected flux densities are given in Tab. \ref{tab:observations_1310}, while the corresponding radio SED in Fig. \ref{fig:1310}. The spectral index is flat from low to high frequencies, ranging from -0.42$\pm$0.06 (0.2-0.88 GHz) to -0.19$\pm$0.12 (4.8-8.6 GHz. As a whole, the unresolved morphology and the flat radio spectrum support the classification as blazar, and in particular as a flat-spectrum radio quasar.

Finally, the \emph{Fermi} $\gamma$-ray spectrum is described by a simple power law with photon index 2.759$\pm$0.135 and a 0.1-100 GeV flux of 7.42$\pm$1.39 $\times$ 10$^{-12}$ erg cm$^{-2}$ s$^{-1}$ \citep{2020ApJ...892..105A}. At the source redshift, this implies a $\gamma$-ray luminosity of $2\times 10^{44}$ erg s$^{-1}$: this luminosity is one of the lowest among \emph{Fermi} flat spectrum radio quasars reported in \cite{2020ApJ...892..105A}. Moreover, the source is reported as variable in the 4FGL catalogue, having a variability index of 74.5, further supporting the blazar classification.


\begin{table}
\caption{Collected radio flux densities for IGR\,J13109--5552.}
\centering
\begin{tabular}{cccccccccccc}
\hline
Telescope   & Frequency         & RMS           & Flux density      \\    
            & (GHz)             & (mJy/beam)    & (mJy)             \\
\hline
GLEAM       & 0.2               & 37            &  778$\pm$60      \\
RACS        & 0.88              & 0.25          &  420$\pm$21      \\   
ATCA        & 4.8               & 2.19          &  280$\pm$14     \\
ATCA        & 8.6               & 3.89          &   250$\pm$13     \\
\hline
\end{tabular}
\label{tab:observations_1310}
\end{table}


\end{document}